# Probing the flat-band limit of the superconducting proximity effect in Twisted Bilayer Graphene Josephson junctions

A. Díez-Carlón[1,2], J. Díez-Mérida[1,2], P. Rout[1,2], D. Sedov[3], P. Virtanen[4], S. Banerjee[3], R. P. S. Penttilä[5], P. Altpeter[1], K. Watanabe[6], T. Taniguchi[7], S.-Y. Yang[8], K. T. Law[9], T. T. Heikkilä[4], P. Törmä[5], M. S. Scheurer[3] and D. K. Efetov[1,2]*

1. Fakultät für Physik, Ludwig-Maximilians-Universität, Schellingstrasse 4, 80799 München, Germany
2. Munich Center for Quantum Science and Technology (MCQST), München, Germany
3. Institute for Theoretical Physics III, University of Stuttgart, 70550 Stuttgart, Germany
4. Department of Physics and Nanoscience Center, University of Jyväskylä, P.O. Box 35 (YFL), FI-40014 University of Jyväskylä, Finland
5. Department of Applied Physics, Aalto University School of Science, FI-00076 Aalto, Finland
6. Research Center for Functional Materials, National Institute for Materials Science, 1-1 Namiki, Tsukuba 305-0044, Japan
7. International Center for Materials Nanoarchitectonics, National Institute for Materials Science, 1-1 Namiki, Tsukuba 305-0044, Japan
8. Southern University of Science and Technology, Shenzhen 518055, P.R. China
9. Department of Physics, Hong Kong University of Science and Technology, Hong Kong, China

*E-mail: dmitri.efetov@lmu.de

**ABSTRACT**

**While extensively studied in normal metals, semimetals and semiconductors, the superconducting (SC) proximity effect remains elusive in the emerging field of flat-band systems. In this study we probe proximity-induced superconductivity in Josephson junctions (JJs) formed between superconducting NbTiN electrodes and twisted bilayer graphene (TBG) weak links. Here the TBG acts as a highly tunable topological flat-band system, which due to its twist-angle dependent bandwidth, allows to probe the SC proximity effect at the crossover from the dispersive to the flat-band limit. Contrary to our original expectations, we find that the SC remains strong even in the flat-band limit, and gives rise to broad, dome shaped SC regions, in the filling dependent phase diagram. In addition, we find that unlike in conventional JJs, the critical current $I_c$ strongly deviates from a scaling with the normal state conductance $G_N$. We attribute these findings to the onset of strong electron interactions, which can give rise to an excess critical current, and also work out the potential importance of quantum geometric terms as well as multiband pairing mechanisms. Our results present the first detailed study of the SC proximity effect in the flat-band limit and shed new light on the mechanisms that drive the formation of SC domes in flat-band systems.**

# I. INTRODUCTION

The superconducting proximity effect at the interface of a superconductor (SC) and a normal metal (N) is understood by the leaking of superconducting correlations into the normal region, through the creation of phase coherent Andreev pairs [1,2]. This process is explained by de Gennes' theory [1–5], which considers wavefunction matching across the N/SC interface, and typically works well for normal metals with dispersive bands, a large Fermi velocity and Fermi surface. However, these conditions are dramatically altered in materials with a vanishing bandwidth, as found in various flat-band systems, like kagome metals [6,7], moiré materials [8] and Lieb lattices [9–11]. Recent theoretical studies have now derived new formalisms to describe the superconducting proximity effect in such systems [12,13]. They show that in atomic flat-bands, the proximity effect can be strongly quenched, as their Fermi velocity vanishes and electrons localize. However strong electronic interactions and quantum geometric terms can give rise to an enhanced conduction and superfluidity [14–16], and as a result allow for a sizeable SC proximity effect [12,13].

Twisted bilayer graphene (TBG) presents a highly suitable platform to test these predictions. It contains two adjacent flat-bands with a bandwidth that can be tuned below $w < 10$ meV in devices with a magic twist-angle of $\theta_m \sim 1.1°$, which is three orders of magnitude lower than in normal metals. These were previously shown to host a broad number of strongly correlated phases, such as superconductivity [17,18], correlated insulators [19,20], non-trivial topological states [21–24], and were argued to contain quantum geometric terms that can enhance the superconducting state [16]. In addition, the close proximity of the two flat bands in energy could in principle allow for the formation of more complex Andreev pairs, that could reside in both bands [25–27].

In this work, we probe the flat-band limit of the SC proximity effect, by performing a detailed study of three twist-angle controlled SC/TBG/SC Josephson junctions (JJs) and their evolution as the bandwidth is tuned from the dispersive limit $w \sim 66$ meV ($\theta \sim 1.24°$), $w \sim 18$ meV ($\theta \sim 0.94°$) to the flat-band limit of $w < 10$ meV ($\theta \sim 1.00°$). Surprisingly we find that, even in devices with the flattest bands, the SC proximity effect can be comparable in strength to the dispersive bands. When the Fermi energy is tuned through the flat-bands, the critical current forms dome shaped regions close to half-filling of the bands, and shows unconventional interference patterns. We discuss our findings in the context of possible contributions to the SC proximity effect through strong electron interactions, quantum geometric terms, multiband pairing processes and provide constrains on its underlying symmetries.

# II. RESULTS

## A. Proximity-induced superconductivity in a TBG JJ

As depicted in Fig. 1a, our devices consist of a van der Waals heterostructure of TBG that is encapsulated with hexagonal boron nitride dielectrics (hBN) and patterned into a rectangular mesa of width $W \sim 1.5$ μm. The devices are capacitively coupled to a $SiO_2$/Si back gate that allows to control the carrier concentration in the TBG, and are contacted with sputtered s-wave superconducting NbTiN leads that form one-dimensional edge contacts, resulting in junctions of length $L \sim 200$ nm. It is worth noting that while several previous TBG JJ experiments used the intrinsic SC state of TBG and a gate-defined weak link [28–30], the here presented device design strongly simplifies the modeling of such JJs, as it uses a SC state with a known pairing mechanism, and defines a simpler and sharper junction interface.

We first focus on device D2 with a twist-angle $\theta \sim 1.00 \pm 0.01°$, slightly smaller than $\theta_m$. All data were obtained at a temperature of 35 mK. Fig. 1b shows two-terminal resistance $R$ measurements as a function of gate voltage $V_g$ as we tune through the dispersive and flat-band regions (see Fig. 1c), which show peaks at integer fillings of the moiré unit cell $v = 0, \pm 2, \pm 4$. This behavior is characteristic of strongly correlated TBG devices in the range of twist-angles of $\theta \sim 1.0°$ - 1.2°, where $v = \pm 2$ mark the occurrence of correlated insulator states [8,17–20].

Additionally, zero-resistance states at various fillings are recorded due to the SC proximity effect. This is seen in Fig. 1f, were we measure the differential resistance $dV/dI$ as a function of d.c. current $I$ for the same range of $V_g$ as Fig. 1b. It clearly shows superconducting regions with zero resistance (dark blue), that are limited by the critical current $I_c$ (displayed in Fig. 1b), which exist across almost the entire density range. We observe broad dome-shaped SC regions in the dispersive bands at $|v| > 4$, and also in the flat-bands near the charge-neutrality point (CNP) $v = 0$, and between fillings $v = \pm 2$ and $v = \pm 4$. We note that the observed $I_c$ in the superconducting domes at $v < -2$ reaches its maximum at $v \sim -2.9$ and spans down to $v \sim -3.5$, exceeding the range of filling where TBG typically shows an intrinsic SC state [8,17,18]. Additionally, we do not observe any transition in the $I_c$ or in the measured interference patterns at these fillings (as we will discuss later in this work). We then conclude that the observed $I_c$ in our JJs is mainly caused by the proximity effect.

The formation of the Josephson effect is confirmed by the observation of non-linear current-voltage characteristics, such as the one shown in Fig. 1d in the dispersive bands ($V_g = 50V$, $v = 7.3$). The switching from the zero-resistance to the normal state is detected as a sharp transition in voltage and presents a hysteretic behavior between the retrapping ($I_r$) and critical ($I_c$) currents, as is common for underdamped junctions or due to self-heating effects [4]. The phase coherence of the JJ is further demonstrated by the observation of Fraunhofer interference patterns when a perpendicular magnetic field $B$ is applied to the junction, as seen in Fig. 1e. The period of the measured oscillations $\Delta B \sim 2.5 \pm 0.2$ mT matches well with the expected periodicity $\Delta B_{phys} \sim 2.3 \pm 0.1$ mT defined by the physical area of the junction $\sim 0.33 \pm 0.07$ $\mu m^2$ when flux-focusing effects are included (see Supplementary B for more details).

**B. Excess of supercurrent due to strong electron interactions in the flat-bands**

The maximal critical currents that we observe in the flat-bands of $I_c \sim 65$ nA are only a factor of 5 lower as compared to the dispersive bands $I_c \sim 350$ nA. This is surprising, as we would have expected a much stronger suppression of the SC proximity effect in the flat-bands as compared to the dispersive bands, given the large reduction in the bandwidth and Fermi velocity [12,13,16]. Since $I_c$ values itself do not allow for a direct estimate of the strength of the SC proximity effect, it is instead typically approximated with the $I_c R_N$ product [2,4,5]. Here $R_N$ is the normal state resistance of the JJ, which we extract through resistance measurements at a current $I > I_c$ (see

Supplementary B for more details). Fig. 2a shows $I_c R_N$ vs. $v$, where we find especially large values in the dome shaped proximity induced regions between fillings $v = \pm 2$ and $v = \pm 4$, that are comparable to the ones in the dispersive bands. This further confirms that unlike the initial expectations, the SC proximity effect is surprisingly large in the flat-bands of TBG [2,5,31].

The unexpected strength of the SC proximity effect in the flat-bands is not the only peculiarity, as also the filling dependence of the $I_c$ inside the flat-bands is highly unusual. In typical JJs, Andreev pairs undergo dephasing processes, the strength of which scales with the normal state resistance $R_N$. It is therefore often found that that $I_c$ correlates with $G_N = R_N^{-1}$, the normal state conductance [2,4,5]. We demonstrate this is the case in JJs with single layer graphene or small twist-angle TBG as the weak links (see Supplementary D for comparison) [32]. The same trend is also found when looking at the dispersive bands of our device D2, as can be seen in Fig. 2b, which shows the filling dependence of the critical current $I_c$ vs. $v$ and overlays it with the filling dependence of the normal state conductance $G_N$ vs. $v$. In this case, the $I_c$ generally follows the density of states by becoming stronger with increasing filling and peaking between $v \sim 7$ and $v \sim 8$ (see Supplementary Fig. S12).

In strong contrast to this, the flat-bands of our devices show the exact opposite trend, as pointed out by vertical dashed arrows in Fig. 2b (see also Fig. 3c). Here both $I_c$ and $G_N$ increase when doping away from the CNP, but while $I_c$ peaks at $|v| \sim 0.3$ and decreases beyond these points, finally vanishing at $|v| \sim 1$, $G_N$ continues to increase until $|v| \sim 1$. At higher doping, where the $I_c$ domes at $|v| > 2$ are, the same conduct appears: because $G_N$ shows smaller values than near the CNP, we expect a vanishing $I_c$ and yet we observe similar ($v > 2$) or even bigger values ($v < -2$). Analogous trends are also found for other devices in their flat-bands (see Fig. 3a-b).

The deviation of the observed scaling of $I_c$ and $G_N$ in the flat-bands could be explained by the existence of excess values of $I_c$ in certain fillings, that could come from an extra contribution that is independent of $G_N$ and thus of band dispersion. Such often-neglected term, $I_c^{\text{int}}$, indeed exists theoretically and scales with an attractive interaction coupling between electrons, contributing to the total critical current because it boosts the Andreev pair transport through the JJ [13]. It is then an important contribution in (quasi-)flat-bands, where the range of pair correlations without interactions can become short due to localization in noninteracting transport. Similarly, as in the superfluid weight, part of this increase is related to the quantum geometry and is independent of band dispersion [12,13,24,33]. As has been shown in Ref. [10] and Supplementary E, $I_c^{\text{int}}$ is independent of $G_N$, so that the total critical current $I_c$ does not necessarily correlate with $G_N$ in regions where $I_c^{\text{int}}$ dominates, as it could be the case for our devices in the ranges mentioned above. We note that, for a quantitative explanation of the observed unconventional relation between $I_c$ and $G_N$, future theoretical models should also derive the full contribution from the band dispersion to the $I_c$ and compare it with $I_c^{\text{int}}$.

### C. Examining potential effects due to quantum geometry and multiband pairing

While we now have postulated the existence of enhanced $I_c$ regions in the flat-bands and a potential explanation related to strong electron interactions, the strong variation of $I_c$ and the formation of dome shaped regions, especially close to the band insulators between fillings $v = \pm 2$ and $v = \pm 4$, remain unclear. In order to get a better understanding of the driving forces behind these domes, we use the powerful tuning knob of TBG, the bandwidth $w$, which can be controlled directly by the twist-angle. As we will see next, by tuning $w$, we can effectively tune the band-dispersion and interactions.

Fig. 3 compares three samples with different twist-angles: D1 ($\theta \sim 0.94 \pm 0.01°$), D2 ($\theta \sim 1.00 \pm 0.01°$) and D3 ($\theta \sim 1.24 \pm 0.01°$). All three devices show a finite $I_c$ across almost the entire flat-band, with some dome shaped regions in-between half-filling $v = \pm 2$ and the band-edges $v = \pm 4$. It appears that as $w$ is lowered (Fig. 4d-f), the dome-shaped SC regions move closer to the band-edges (see Fig. 3a-c). While $I_c$ is suppressed in D1 and D3 close to the CNP, in D2, which has the lowest bandwidth, the region around the CNP shows enhanced $I_c$ values.

We try to qualitatively understand the observed variation of $I_c$ with filling and twist-angle. We use the Eliashberg formalism and perturbation theory to compute the linear response function to the contact-induced pairing from an s-wave superconductor into TBG at the Fermi level $E_F$. Under the slowly varying field approximation, the response is captured by the superconducting pair correlation function $\phi_R = \int_{u.c.} da \, \langle c_{\downarrow,-}(R+a) \cdot c_{\uparrow,+}(R+a) \rangle$. To model the TBG we use the Bistritzer-MacDonald continuum model (BM) [34], and calculate the band-structure, bandwidth (see Fig. 4d-f) and the pair correlator $\phi_R$ at a distance $R$ in moiré unit cells coinciding with the center of the junctions D1-3 (see Supplementary F for more details). Here, besides the dominant induced Andreev pair components with opposite spin $(\uparrow, \downarrow)$ and valleys $(-, +)$, a possible small admixed intravalley component in TBG is expected to be quickly suppressed with $R$. Since the correlator $\phi_R$ can be understood as a response function of the TBG bands to an external superconducting field, a qualitative comparison can be made with $I_c$.

The strength of this formalism lies in that we can explicitly account for the different contributions to the correlator $\phi_R$, which include the band dispersion and the quantum geometric terms. Importantly, the induced Andreev pairs do not need to reside only in one single band at the Fermi level $E_F$, but can also have weight across multiple bands. The contribution from multiband pairing to $\phi_R$, however, diminishes the further these bands are from $E_F$ and the larger their dispersion is, and thus is expected to be significant only in the flattest bands (see Supplementary F for more details).

In order to highlight the effect of each contribution and work out the effects of the quantum geometry and multi-band pairing, we separate these terms out and plot them independently in Fig. 4. Here we consider three cases. First $\phi_R^{S.B.}$, which contains only dispersion-driven effects by taking the atomic limit of the flat-bands, where the Bloch states are completely momentum independent and the quantum geometric contribution is suppressed, and Andreev pairs that are only formed inside one single band at $E_F$ (Fig. 4a). We second consider $\phi_R^{Q.G.}$, which contains dispersion-driven effects but now departing from the atomic limit where we incorporate the overlap between Bloch states at different momenta, which is determined by the quantum metric of the bands, and Andreev pairs that are only formed inside one band at $E_F$ (Fig. 4b). Lastly, we consider $\phi_R^{M.B.}$, which contains dispersion-driven effects with a suppressed quantum geometric contribution, like in the first case, but now additional Andreev pairs formed inside other bands can interfere with the ones from the band at $E_F$ (Fig. 4c).

Our calculations are shown in Fig. 4g-i for the three different bandwidths $w$ corresponding to the three devices shown in Fig. 3, with D3 ($w \sim 66$ meV), D1 ($w \sim 18$ meV) and D2 ($w \sim 4$ meV) (BM band-structures shown in Fig. 4d-f). Here we find that, when the filling is varied, $\phi_R^{S.B.}$ just follows the density of states for all $w$ (Supplementary Fig. S11 shows this statement explicitly). It shows peaks close to half-filling $v = \pm 2$, which rapidly decrease close to the band edges $v = \pm 4$ and the CNP. Although this scenario very well matches the broadest bandwidth $w \sim 66$ meV (Fig. 4g) with our $I_c$ measurements (Fig. 3a), it dramatically fails to describe the observed features of devices D1 and D2 with lower $w$ (see Fig. 4h-i and Fig. 3b-c). For the flattest bands, $w \sim 4$ meV, the peaks in $\phi_R^{S.B.}$ strongly broaden and show an almost constant value across the entire range of $v$ (Fig. 4i), failing to produce dome shaped regions. By now plotting $\phi_R^{Q.G.}$ including quantum geometric and $\phi_R^{M.B.}$ including multiband terms, we find that these are strongly altered from $\phi_R^{S.B.}$ as the bandwidth is lowered, and both give rise to dome shaped features close to the band-edges, just as observed for the devices D1 with $w \sim 18$ meV and D2 with $w \sim 4$ meV. We find that overall, the $\phi_R^{Q.G.}$ term better matches the findings of the $w \sim 4$ meV case, as it allows for a finite induced-superconducting phase in the center of the band and close to the CNP, as observed in device D2. We also find that D1 with $w \sim 18$ meV is better matched with $\phi_R^{M.B.}$. The moderate mismatch in D2 compared to D1 and D3 could come from the fact that only the former shows correlated states due to interactions, which are not considered in our model.

We want to stress that to fully capture and explain the proximity effect in this correlated system, future theoretical models should further account for interactions, as our current approach does not explain the vanishing $I_c$ due to the correlated states at integer fillings. Overall and despite the simplicity of the continuum model in describing the flat-bands of TBG, the qualitative agreement with the data highlights that when the bandwidth reaches the flat-band limit, quantum geometric and multiband processes could become important in the understanding the susceptibility of TBG to develop superconducting phases [26]. It may explain the formation of dome shaped SC regions between half-filling and the band edges, which roughly coincide with the regions that typically also show intrinsic superconductivity in TBG devices at the magic angle. It is therefore interesting to consider whether similar effects, as the ones that were worked out here, can also explain the position of the SC domes in TBG.

### D. Symmetry breaking and the Josephson diode effect

The presence of interactions in the TBG JJs at half-fillings of the bands is further suggested by a consistent observation of a symmetry-broken Josephson effect, which will be the focus of the remainder of this work. We start by studying an interference pattern in D2 at $v = -2.5$, shown in Fig. 5a. When recording $I_c$ vs. $B$ in opposite directions of the d.c. current, $I_c^+(B)$ and $I_c^-(B)$, we find that $I_c^+(B) \neq |I_c^-(B)|$, as clearly seen in the mid panel of Fig. 5a. Such observation indicates inversion symmetry is broken in our JJs and is unlike the conventional symmetric patterns found near the CNP or in the dispersive bands (see Fig. 1e and Supplementary B). This non-reciprocity is a hallmark of the Josephson diode effect (JDE), which requires both inversion, $C_{2z}$, and time-reversal symmetry (TRS) breaking [35–37]. Importantly, in our devices TRS is broken by applying an external perpendicular magnetic field to the graphene layers. This rules out Rashba spin-orbit coupling as a possible mechanism, where an in-plane magnetic field that is perpendicular to the direction of the current is needed to produce the non-reciprocity [37]. Furthermore, the TBG weak link itself does not intrinsically break TRS, as no asymmetry is recorded at zero field, i.e. we find $I_c^+(0) = |I_c^-(0)|$. This is further confirmed by having $I_c^+(B) = |I_c^-(-B)|$ (see bottom panel of Fig. 5a); an expression that conserves this said symmetry and that makes the diode programmable by applying exact opposite fields. Such programmability is illustrated in Fig. 5b-c, where the d$V$/d$I$ curves validate the aforementioned symmetry relations and the operation of the diode is demonstrated in the rectification measurements.

Notably, we find that the key features reported in Fig. 5a-c extend across the entire filling of the dome, spanning from $v \sim$ -2 to $v \sim$ -3.5. This is shown in Fig. 5d, where the diode efficiency parameter $\eta(B) = (I_c^+(B) - |I_c^-(B)|)/(I_c^+(B) + |I_c^-(B)|)$, calculated at half the superconducting magnetic flux quantum $\Phi_0/2$, is found to correlate with the $I_c$ of the dome. This shows that the asymmetry is most pronounced at the center of the dome at $v$ = -2.9, but is no longer detectable at the edges of it at $v$ = -2.3 and $v$ = -3.4. Furthermore, our observation of inversion symmetry breaking consistently appears in the domes at the hole side of $v \sim$ -2 for all our TBG JJs close to $\theta_m$ (D1-3) and, in some cases (D1-2), on the electron side of $v \sim$ 2. Fig. 5e shows this, where for all devices $\eta$ correlates with the $I_c$ of their respective domes. In this case, to compare the JDE between different devices we have calculated the maximum value of $\eta$ between $-2\Phi_0$ and $2\Phi_0$ (see Supplementary K for more details). Such extent of the asymmetry with filling suggests a distinct phase to be responsible for the JDE.

Several interacting ground states that spontaneously break the $C_{2z}$ and spinless time-reversal symmetries of TBG have been proposed. Valley polarization was suggested to explain the abundance of orbital magnetism and broken inversion found at these bands [30,38–40], although original nematicity measurements [41,42] showed $C_{3z}$ symmetry-breaking instead. In addition, recent experiments in scanning tunneling microscopy have pointed towards the most likely candidates at such fillings having intervalley coherent or incommensurate Kekulé spiral orders [43], which do not break $C_{2z}$. Therefore, none of the above candidates are consistent with our findings, and a sublattice-polarized phase emerges as the only candidate that fulfills our observed symmetry relations [26,44,45], where each valley carries opposite Chern numbers $C = 1$ and $C = -1$.

Given the range of twist-angles and fillings where the JDE is found, a natural question is whether the presence of the intrinsic superconducting phase of TBG could play a significant role. However, the observation of oscillations in $I_c$, with their period matching the total area of the junction (Fig. 5a), argues against the TBG being fully intrinsically superconducting in these devices [29,46,47]. The independence in temperature of the inductance associated with the junction is also consistent with this (see Supplementary J), in contrast with previous observations of asymmetric oscillations in SQUIDs with high kinetic inductance [48].

Finally, the inversion symmetry breaking resulting from geometrical factors, such as a non-uniform junction with different width of contacts, can be ruled out in our case, given that this effect would be independent of the electron density and yet, the JDE is only observed in the flat-bands and in devices close to $\theta_m$ (see Supplementary D). Self-field effects caused by inhomogeneous current bias and screening currents [2,49] cannot be the cause of the JDE either, since our small $I_c$ results in a bigger Josephson penetration length $\lambda_J = \Phi_0 tW/4\pi\mu_0 I_c \lambda_L^2 \sim 7$ μm compared to the dimensions of the junction [2,50,51]. Here $t \sim 0.6$ nm is the thickness of TBG and $\lambda_L \sim 400$nm is the London penetration length of NbTiN [52]. In Fig. 5a it can be observed that the nodes of the $I_c$ are lifted, which is a signature of an asymmetry in the supercurrent density profile [53,54]. Nevertheless, this node-lifting effect is in general not enough to provide a JDE though, and on top of that higher-order terms that bring the current-phase relation into a non-sinusoidal form are needed [37,55]. Such higher harmonics or anomalous phases could be a result of a symmetry-broken state as discussed above [56,57], or by topological edge states [58,59]. The latter could also be related to our samples, given that the supercurrent is mostly carried by the edges where the JDE is observed (see Supplementary L). Testing these possible mechanisms would require phase sensitive measurements, which we leave for future works.

## III. CONCLUSIONS

To summarize, we have explored the versatility of TBG to support an extrinsic Josephson effect in its entire electronic band structure. The dispersive bands, as well as other studied devices away from the magic angle, served us to compare the proximity effect with the flat-band limit and test its prediction. Our observations suggest an excess critical current $I_c$ and the formation of SC domes in the flat-band limit could hint at the importance of electron interactions and may be enhanced by quantum geometric and multiband pairing effects. Future works should be focused on incorporating strong interactions in the mechanism of induced superconductivity in this system, in order to explain symmetry-broken effects such as the Josephson diode effect.

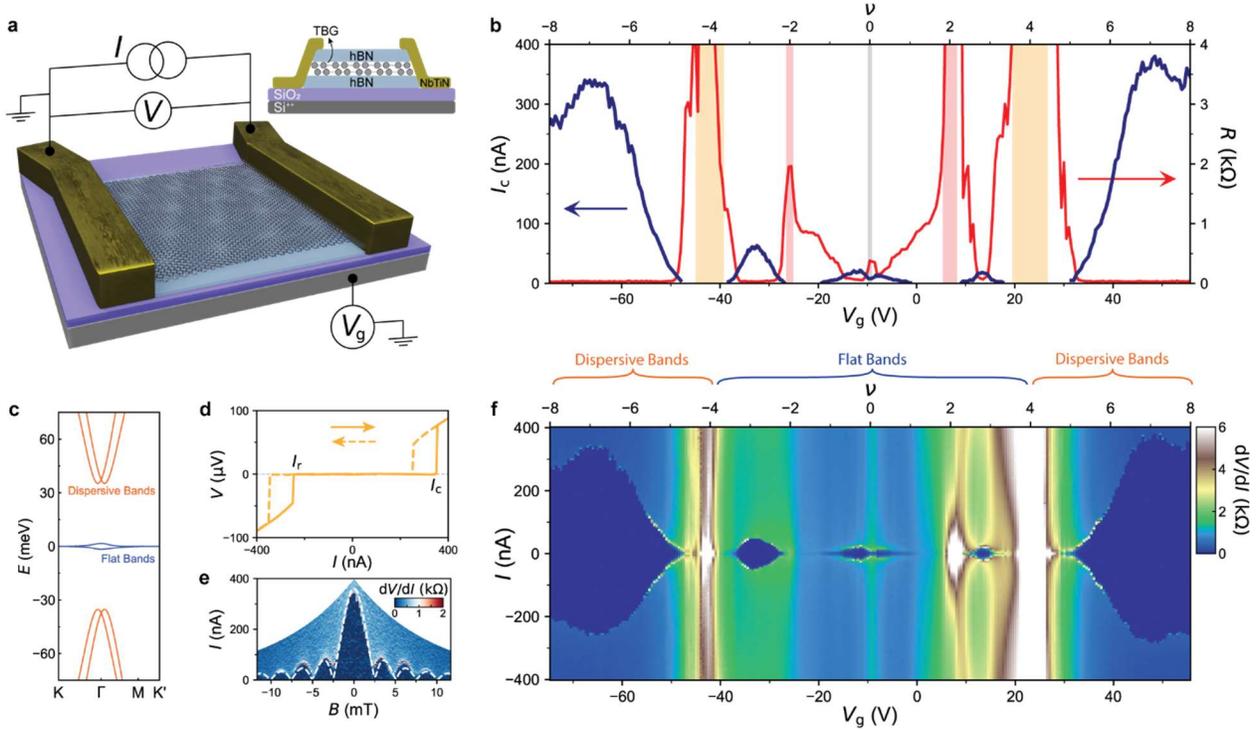

**Fig. 1. Superconducting proximity effect in a TBG Josephson junction. a,** Device schematic of a TBG sheet acting as the weak link of a JJ. The voltage $V$ across the junction is recorded as a current bias $I$ is applied through the superconducting electrodes in a two-probe measurement. The carrier density is tuned by a gate voltage $V_g$ to the doped Si. **b,** Resistance $R$ in red (right axis) at zero current bias, as a function of $V_g$ (bottom) and of the corresponding moiré filling factor $\nu$ (top). Shaded vertical lines indicate the presence of the charge-neutrality point (grey), the correlated insulators at half-filling of the flat-bands (red), and the band insulators between the flat and dispersive bands (yellow). Regions with low resistance have a finite critical current $I_c$ (blue, left axis), extracted from the non-linear characteristics measured in **f**. **c,** Band structure of TBG for $\theta = 1.00°$. **d,** IV curve measured at $V_g = 50$ V. The solid and dashed lines have opposite sweep directions as indicated by the arrows. **e,** Interference pattern recorded at the dispersive bands, for $V_g = 60$ V, which agrees well with a uniform single-slit junction (white dashed line). **f,** Differential resistance d$V$/d$I$ map, where dark blue regions represent superconducting states. Their contour along positive values of $I$ was used for extracting $I_c$ in **b**. All data was obtained in device D2.

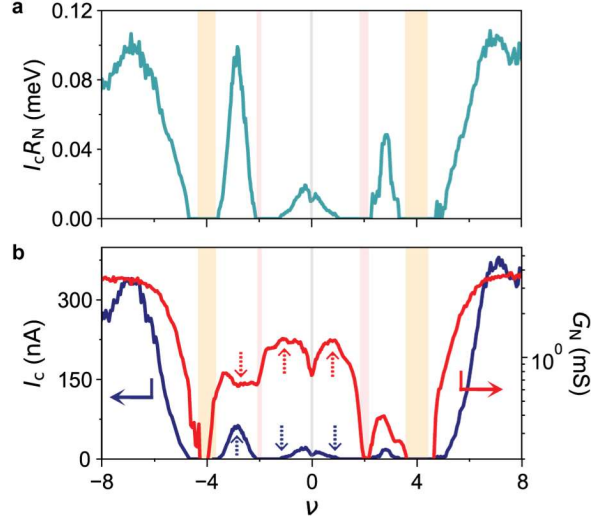

**Fig. 2. Strength of the proximity effect and relation between critical current and normal conductance. a,** Product of the critical current and normal-state resistance, $I_cR_N$, as a function of the moiré filling factor $\nu$. **b,** Critical current $I_c$ in blue (left axis) and normal-state conductance $G_N$ in red (right axis), both as a function of $\nu$. Following the same color code, the dashed vertical arrows indicate whether the corresponding quantity has reached a maximum or minimum. This highlights the unusual relation between $I_c$ and $G_N$ found at the flat-bands. All data corresponds to D2. Shaded vertical lines indicate the presence of the charge-neutrality point (gray), the correlated insulators (red), and the band insulators (yellow).

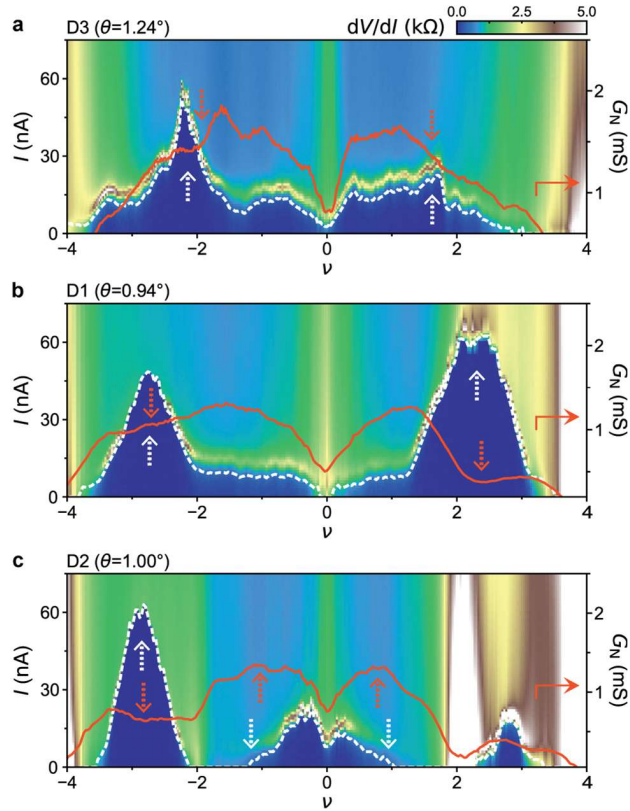

**Fig. 3. Proximitized flat-bands with varying bandwidth by tuning the twist-angle. a-c,** Colormaps of measured differential resistance $dV/dI$ as a function of d.c. current $I$ (left axis) and filling $\nu$ of the flat-band, for devices D3, D1 and D2, respectively. The solid red line corresponds to the normal state conductance $G_N$ (right axis). The critical current $I_c$ of each device is extracted by following the contour of the dark-blue regions in the colormaps, marked by white-dashed lines. Following the same color code, the dashed vertical arrows indicate the regions where $I_c$ and $G_N$ follow opposite trends, pointing whether the corresponding quantity is an increasing or decreasing function of $|\nu|$. In D1 and D3, the observed Josephson effect holds throughout almost the entirety of the flat-bands, given the absence of correlated states at integer fillings. In the case of D2 (the closest to $\theta_m$), the superconducting phases get interrupted by insulating states at half-filling.

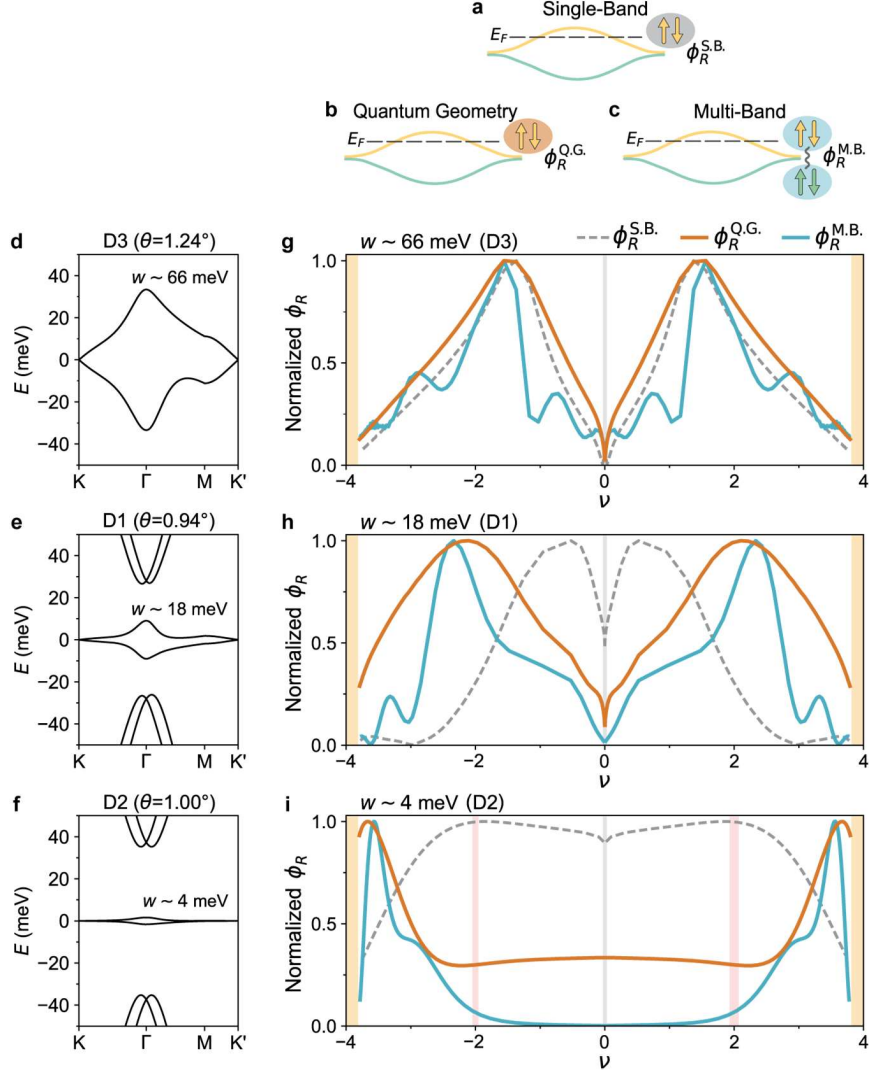

**Fig. 4**. **Quantum geometric and multiband contributions to the induced superconductivity in the flat-bands. a-c,** Sketches illustrating the different contributions to the induced superconductivity. The color of the spin of the electrons forming the Andreev pair indicates which band they are coming from. In **a**, the contribution comes only from the dispersion of the same band where the Fermi level $E_F$ lies, whereas in **c** the interference with more bands is also accounted for, i.e. multiband pairing. In **b**, the quantum geometric contribution is considered along with the dispersion of the single-band in **a**. **d-f,** Band structure of the TBG continuum model along high-symmetry points for the three different twist-angles corresponding to samples D3, D1 and D2, respectively. The bandwidth $w$ of the flat-bands is also shown. **g-i,** Computed superconducting correlator $\phi_R$ vs. filling factor $v$ for the three processes illustrated in **a-c**, following the same color-code as the sketches of the Andreev pairs. Shaded vertical lines indicate the presence of the charge-neutrality point (gray), the correlated insulators (red), and the band insulators (yellow).

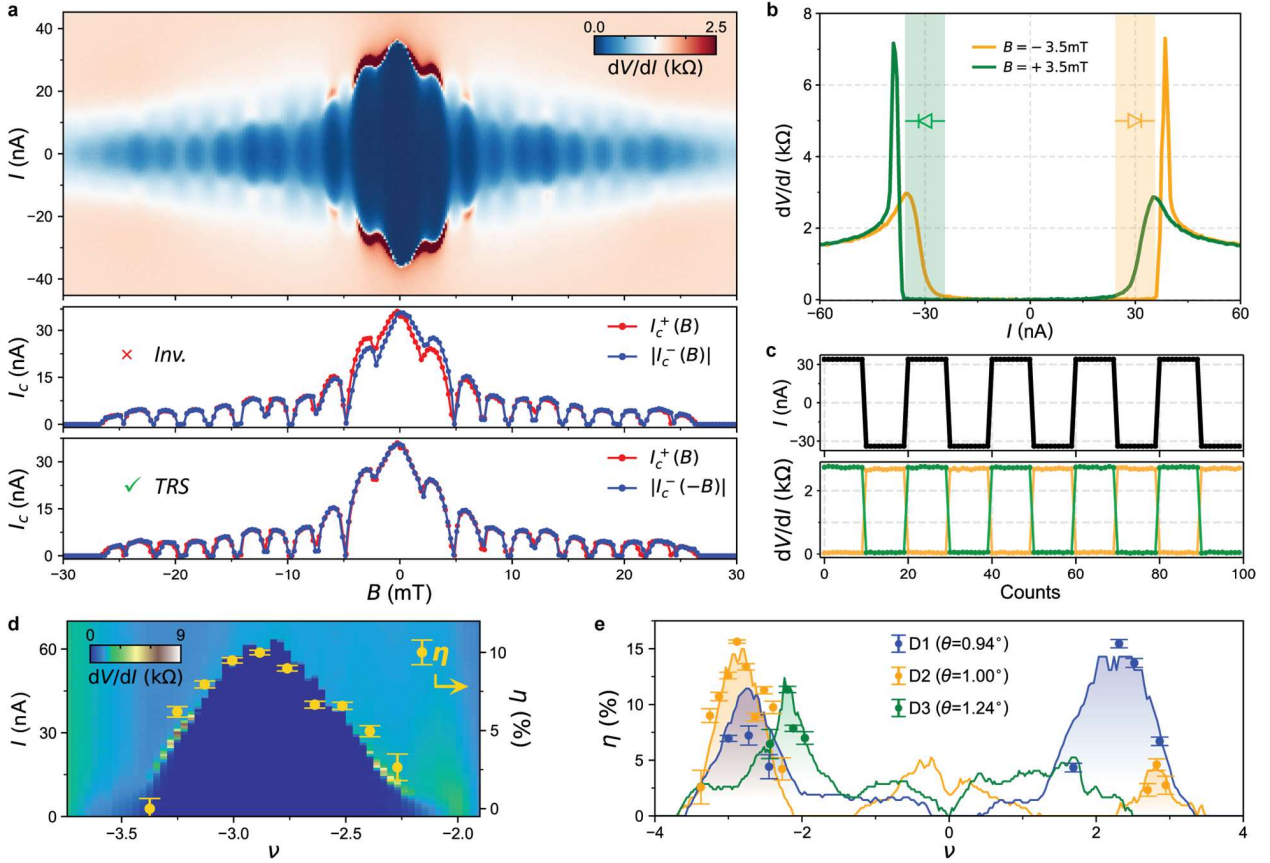

**Fig. 5. Josephson diode effect and inversion symmetry breaking at the |ν| > 2 domes. a,** Top panel shows measured differential resistance d$V$/d$I$ as a function of d.c. bias current $I$ and magnetic field $B$, at $v$ = -2.5 for D2. The critical current of the dark-colored oscillations is extracted for positive ($I_c^+$) and negative ($I_c^-$) directions of the d.c. current. These are represented for the same and opposite values of magnetic field in the middle and bottom panels, respectively. **b,** d$V$/d$I$ traces measured at opposite magnetic fields. Shaded regions mark the $I$ values at which the Josephson diode is operational. The measurements are performed at $v$ = -2.9 for D2. **c,** Demonstration of the reversible JDE in **b**, performed by switching between the superconducting and normal states when opposite currents are applied, in this case +34 nA and -34 nA. Reversibility of the direction of the diode is achieved by applying an exact opposite $B$. **d,** d$V$/d$I$ colormap vs. $I$ (left axis) and $v$. Represented along with error bars is the extracted diode efficiency $\eta$ (right axis) vs. $v$, computed at a magnetic flux $\Phi = \Phi_0/2$. All data corresponds to D2. **e,** Maximum value of $\eta(B)$ between $-2\Phi_0$ and $2\Phi_0$, represented with error bars and as a function of $v$ for all junctions near the magic angle D1-3. The line plots correspond to the $I_c$ (in arbitrary units) from Fig. 3d-f. The shaded regions correspond to fillings where a finite asymmetry was recorded. The interference patterns from where $\eta$ was extracted can be found in Supplementary I.

**References:**

[1] P. G. De Gennes, Boundary Effects in Superconductors, Rev. Mod. Phys. **36**, 225 (1964).

[2] K. K. Likharev, Superconducting weak links, Rev. Mod. Phys. **51**, 101 (1979).

[3] T. M. Klapwijk, Proximity Effect From an Andreev Perspective, J. Supercond. **17**, 593 (2004).

[4] M. Tinkham, *Introduction to Superconductivity* (Dover, Mineola, 1996).

[5] A. A. Golubov, M. Yu. Kupriyanov, and E. Il'ichev, The current-phase relation in Josephson junctions, Rev. Mod. Phys. **76**, 411 (2004).

[6] Z. Li et al., Realization of flat band with possible nontrivial topology in electronic Kagome lattice, Sci. Adv. **4**, eaau4511 (2018).

[7] M. Kang et al., Dirac fermions and flat bands in the ideal kagome metal FeSn, Nat. Mater. **19**, 163 (2020).

[8] L. Balents, C. R. Dean, D. K. Efetov, and A. F. Young, Superconductivity and strong correlations in moiré flat bands, Nat. Phys. **16**, 725 (2020).

[9] S. Mukherjee, A. Spracklen, D. Choudhury, N. Goldman, P. Öhberg, E. Andersson, and R. R. Thomson, Observation of a Localized Flat-Band State in a Photonic Lieb Lattice, Phys. Rev. Lett. **114**, 245504 (2015).

[10] A. Julku, S. Peotta, T. I. Vanhala, D.-H. Kim, and P. Törmä, Geometric Origin of Superfluidity in the Lieb-Lattice Flat Band, Phys. Rev. Lett. **117**, 045303 (2016).

[11] S. Ahmadkhani and M. V. Hosseini, Superconducting proximity effect in flat band systems, J. Phys. Condens. Matter **32**, 315504 (2020).

[12] Z. C. F. Li, Y. Deng, S. A. Chen, D. K. Efetov, and K. T. Law, *Flat Band Josephson Junctions with Quantum Metric*, arXiv:2404.09211.

[13] P. Virtanen, R. P. S. Penttilä, P. Törmä, A. Díez-Carlón, D. K. Efetov, and T. T. Heikkilä, *Superconducting Junctions with Flat Bands*, arXiv:2410.23121.

[14] S. Peotta and P. Törmä, Superfluidity in topologically nontrivial flat bands, Nat. Commun. **6**, 1 (2015).

[15] P. Törmä, S. Peotta, and B. A. Bernevig, Superfluidity and Quantum Geometry in Twisted Multilayer Systems, Nat. Rev. Phys. **4**, 528 (2022).

[16] H. Tian et al., Evidence for Dirac flat band superconductivity enabled by quantum geometry, Nature **614**, 440 (2023).

[17] Y. Cao, V. Fatemi, S. Fang, K. Watanabe, T. Taniguchi, E. Kaxiras, and P. Jarillo-Herrero, Unconventional superconductivity in magic-angle graphene superlattices, Nature **556**, 43 (2018).

[18] X. Lu et al., Superconductors, orbital magnets and correlated states in magic-angle bilayer graphene, Nature **574**, 653 (2019).

[19] Y. Cao et al., Correlated insulator behaviour at half-filling in magic-angle graphene superlattices, Nature **556**, 80 (2018).

[20] M. Yankowitz, S. Chen, H. Polshyn, Y. Zhang, K. Watanabe, T. Taniguchi, D. Graf, A. F. Young, and C. R. Dean, Tuning superconductivity in twisted bilayer graphene, Science **363**, 1059 (2019).

[21] M. Serlin, C. L. Tschirhart, H. Polshyn, Y. Zhang, J. Zhu, K. Watanabe, T. Taniguchi, L. Balents, and A. F. Young, Intrinsic quantized anomalous Hall effect in a moiré heterostructure, Science **367**, 900 (2020).

[22] K. P. Nuckolls, M. Oh, D. Wong, B. Lian, K. Watanabe, T. Taniguchi, B. A. Bernevig, and A. Yazdani, Strongly correlated Chern insulators in magic-angle twisted bilayer graphene, Nat. 2020 5887839 **588**, 610 (2020).

[23] I. Das, X. Lu, J. Herzog-Arbeitman, Z.-D. Song, K. Watanabe, T. Taniguchi, B. A. Bernevig, and D. K. Efetov, Symmetry-broken Chern insulators and Rashba-like Landau-level crossings in magic-angle bilayer graphene, Nat. Phys. **17**, 710 (2021).

[24] S. A. Chen and K. T. Law, Ginzburg-Landau Theory of Flat-Band Superconductors with Quantum Metric, Phys. Rev. Lett. **132**, 026002 (2024).

[25] A. Bussmann-Holder, H. Keller, A. Simon, and A. Bianconi, Multi-Band Superconductivity and the Steep Band/Flat Band Scenario, Condens. Matter **4**, 4 (2019).

[26] M. Christos, S. Sachdev, and M. S. Scheurer, Nodal band-off-diagonal superconductivity in twisted graphene superlattices, Nat. Commun. **14**, 1 (2023).

[27] R. Khasanov, B.-B. Ruan, Y.-Q. Shi, G.-F. Chen, H. Luetkens, Z.-A. Ren, and Z. Guguchia, Tuning of the flat band and its impact on superconductivity in $Mo_5Si_3−xP_x$, Nat. Commun. **15**, 2197 (2024).

[28] F. K. de Vries, E. Portolés, G. Zheng, T. Taniguchi, K. Watanabe, T. Ihn, K. Ensslin, and P. Rickhaus, Gate-defined Josephson junctions in magic-angle twisted bilayer graphene, Nat. Nanotechnol. **16**, 7 (2021).

[29] D. Rodan-Legrain, Y. Cao, J. M. Park, S. C. de la Barrera, M. T. Randeria, K. Watanabe, T. Taniguchi, and P. Jarillo-Herrero, Highly tunable junctions and non-local Josephson effect in magic-angle graphene tunnelling devices, Nat. Nanotechnol. **16**, 7 (2021).

[30] J. Díez-Mérida et al., Symmetry-broken Josephson junctions and superconducting diodes in magic-angle twisted bilayer graphene, Nat. Commun. **14**, 1 (2023).

[31] P. Dubos, H. Courtois, B. Pannetier, F. K. Wilhelm, A. D. Zaikin, and G. Schön, Josephson critical current in a long mesoscopic S-N-S junction, Phys. Rev. B **63**, 064502 (2001).

[32] H. B. Heersche, P. Jarillo-Herrero, J. B. Oostinga, L. M. K. Vandersypen, and A. F. Morpurgo, Bipolar supercurrent in graphene, Nature **446**, 56 (2007).

[33] J.-X. Hu, S. A. Chen, and K. T. Law, *Anomalous Coherence Length in Superconductors with Quantum Metric*, arXiv:2308.05686.

[34] R. Bistritzer and A. H. MacDonald, Moiré bands in twisted double-layer graphene, Proc. Natl. Acad. Sci. U. S. A. **108**, 12233 (2011).

[35] J. Hu, C. Wu, and X. Dai, Proposed Design of a Josephson Diode, Phys. Rev. Lett. **99**, 067004 (2007).

[36] F. Ando, Y. Miyasaka, T. Li, J. Ishizuka, T. Arakawa, Y. Shiota, T. Moriyama, Y. Yanase, and T. Ono, Observation of superconducting diode effect, Nat. 2020 5847821 **584**, 373 (2020).

[37] M. Nadeem, M. S. Fuhrer, and X. Wang, The superconducting diode effect, Nat. Rev. Phys. **5**, 558 (2023).

[38] J.-X. Lin, P. Siriviboon, H. D. Scammell, S. Liu, D. Rhodes, K. Watanabe, T. Taniguchi, J. Hone, M. S. Scheurer, and J. I. A. Li, Zero-field superconducting diode effect in small-twist-angle trilayer graphene, Nat. Phys. **18**, 10 (2022).

[39] C.-C. Tseng, X. Ma, Z. Liu, K. Watanabe, T. Taniguchi, J.-H. Chu, and M. Yankowitz, Anomalous Hall effect at half filling in twisted bilayer graphene, Nat. Phys. **18**, 1038 (2022).

[40] J.-X. Hu, Z.-T. Sun, Y.-M. Xie, and K. T. Law, Josephson Diode Effect Induced by Valley Polarization in Twisted Bilayer Graphene, Phys. Rev. Lett. **130**, 266003 (2023).

[41] Y. Jiang, X. Lai, K. Watanabe, T. Taniguchi, K. Haule, J. Mao, and E. Y. Andrei, Charge order and broken rotational symmetry in magic-angle twisted bilayer graphene, Nature **573**, 91 (2019).

[42] Y. Cao, D. Rodan-Legrain, J. M. Park, N. F. Q. Yuan, K. Watanabe, T. Taniguchi, R. M. Fernandes, L. Fu, and P. Jarillo-Herrero, Nematicity and competing orders in superconducting magic-angle graphene, Science **372**, 264 (2021).

[43] K. P. Nuckolls et al., Quantum textures of the many-body wavefunctions in magic-angle graphene, Nature **620**, 525 (2023).

[44] N. Bultinck, E. Khalaf, S. Liu, S. Chatterjee, A. Vishwanath, and M. P. Zaletel, Ground State and Hidden Symmetry of Magic-Angle Graphene at Even Integer Filling, Phys. Rev. X **10**, 031034 (2020).

[45] M. Christos, S. Sachdev, and M. S. Scheurer, Correlated Insulators, Semimetals, and Superconductivity in Twisted Trilayer Graphene, Phys. Rev. X **12**, 021018 (2022).

[46] R. Jha, M. Endres, K. Watanabe, T. Taniguchi, M. Banerjee, C. Schönenberger, and P. Karnatak, *Large Tunable Kinetic Inductance in a Twisted Graphene Superconductor*, arXiv:2403.02320.

[47] M. R. Sinko, S. C. De La Barrera, O. Lanes, K. Watanabe, T. Taniguchi, S. Tan, D. Pekker, M. Hatridge, and B. M. Hunt, Superconducting contact and quantum interference between two-dimensional van der Waals and three-dimensional conventional superconductors, Phys. Rev. Mater. **5**, 014001 (2021).

[48] E. Portolés, S. Iwakiri, G. Zheng, P. Rickhaus, T. Taniguchi, K. Watanabe, T. Ihn, K. Ensslin, and F. K. de Vries, A tunable monolithic SQUID in twisted bilayer graphene, Nat. Nanotechnol. 2022 1711 **17**, 1159 (2022).

[49] T. Golod and V. M. Krasnov, Demonstration of a superconducting diode-with-memory, operational at zero magnetic field with switchable nonreciprocity, Nat. Commun. **13**, 1 (2022).

[50] F. Tafuri, *Fundamentals and Frontiers of the Josephson Effect*, Vol. 286 (Springer International Publishing, 2019).

[51] J. R. Clem, Josephson junctions in thin and narrow rectangular superconducting strips, Phys. Rev. B - Condens. Matter Mater. Phys. **81**, (2010).

[52] J. G. Kroll et al., Magnetic-Field-Resilient Superconducting Coplanar-Waveguide Resonators for Hybrid Circuit Quantum Electrodynamics Experiments, Phys. Rev. Appl. **11**, 064053 (2019).

[53] R. C. Dynes and T. A. Fulton, Supercurrent Density Distribution in Josephson Junctions, Phys. Rev. B **3**, 3015 (1971).

[54] S. Hart, H. Ren, T. Wagner, P. Leubner, M. Mühlbauer, C. Brüne, H. Buhmann, L. W. Molenkamp, and A. Yacoby, Induced superconductivity in the quantum spin Hall edge, Nat. Phys. **10**, 638 (2014).

[55] S. Chen et al., Current induced hidden states in Josephson junctions, Nat. Commun. **15**, 8059 (2024).

[56] M. Alvarado, P. Burset, and A. L. Yeyati, Intrinsic nonmagnetic $\phi_{0}$ Josephson junctions in twisted bilayer graphene, Phys. Rev. Res. **5**, L032033 (2023).

[57] H. Sainz-Cruz, P. A. Pantaleón, V. T. Phong, A. Jimeno-Pozo, and F. Guinea, Junctions and Superconducting Symmetry in Twisted Bilayer Graphene, Phys. Rev. Lett. **131**, 016003 (2023).

[58] C.-Z. Chen, J. J. He, M. N. Ali, G.-H. Lee, K. C. Fong, and K. T. Law, Asymmetric Josephson effect in inversion symmetry breaking topological materials, Phys. Rev. B **98**, 075430 (2018).

[59] M. Endres, A. Kononov, H. S. Arachchige, J. Yan, D. Mandrus, K. Watanabe, T. Taniguchi, and C. Schönenberger, Current–Phase Relation of a WTe2 Josephson Junction, Nano Lett. **23**, 4654 (2023).

**Acknowledgements:**

With thank Srijit Goswami for help in sample fabrication. D.K.E. acknowledges funding from the European Research Council (ERC) under the European Union's Horizon 2020 research and innovation program (grant agreement No. 852927), the German Research Foundation (DFG) under the priority program SPP2244 (project No. 535146365), the EU EIC Pathfinder Grant "FLATS" (grant agreement No. 101099139) and the Keele, Kavli, Tschira and Wells Foundations as part of the SuperC collaboration. K.W. and T.T. acknowledge support from the Elemental Strategy Initiative conducted by the MEXT, Japan (grant number JPMXP0112101001) and JSPS KAKENHI (grant numbers 19H05790, 20H00354, and 21H05233). R.P.S.P. acknowledges financial support from the Fortum and Neste Foundation. This work was supported by the Research Council of Finland under project numbers 339313 and 354735, by European Union's HORIZON-RIA programme 331 (Grant Agreement No. 101135240 JOGATE), by Jane and Aatos Erkko Foundation, Keele Foundation, and Magnus Ehrnrooth Foundation as part of the SuperC collaboration, and by a grant from the Simons Foundation (SFI-MPS-NFS-00006741-12, P.T.) in the Simons Collaboration on New Frontiers in Superconductivity. D.S., S.B., and M.S.S. acknowledge funding by the European Union (ERC-2021-STG, Project 101040651---SuperCorr). Views and opinions expressed are however those of the authors only and do not necessarily reflect those of the European Union or the European Research Council Executive Agency. Neither the European Union nor the granting authority can be held responsible for them.

**Author contributions:**

A.D.C., S.Y.Y. and D.K.E. conceived and designed the experiments; A.D.C. and P.R. fabricated the devices; A.D.C. performed the measurements and analyzed the data; D.S., P.V., S.B., R.P.S.P., T.T.H., P.T. and M.S.S. performed the theoretical analysis; T.T. and K.W. provided materials; S.G. and D.K.E. supported the experiments; A.D.C. and D.K.E. wrote the paper with input from J.D.M., D.S., P.V., T.T.H., P.T., and M.S.S.

**Competing interests:**

The authors declare no competing interests.

**Data availability:**

The data that support the findings of this study are available from the corresponding author upon reasonable request.

# Supplementary Material: Probing the flat-band limit of the superconducting proximity effect in Twisted Bilayer Graphene Josephson junctions

A. Díez-Carlón[1,2], J. Díez-Mérida[1,2], P. Rout[1,2], D. Sedov[3], P. Virtanen[4], S. Banerjee[3], R. P. S. Penttilä[5], P. Altpeter[1,2], K. Watanabe[6], T. Taniguchi[7], S.-Y. Yang[8], K. T. Law[9], T. T. Heikkilä[4], P. Törmä[5], M. S. Scheurer[3] and D. K. Efetov[1,2]*

1. Fakultät für Physik, Ludwig-Maximilians-Universität, Schellingstrasse 4, 80799 München, Germany
2. Munich Center for Quantum Science and Technology (MCQST), München, Germany
3. Institute for Theoretical Physics III, University of Stuttgart, 70550 Stuttgart, Germany
4. Department of Physics and Nanoscience Center, University of Jyväskylä, P.O. Box 35 (YFL), FI-40014 University of Jyväskylä, Finland
5. Department of Applied Physics, Aalto University School of Science, FI-00076 Aalto, Finland
6. Research Center for Functional Materials, National Institute for Materials Science, 1-1 Namiki, Tsukuba 305-0044, Japan
7. International Center for Materials Nanoarchitectonics, National Institute for Materials Science, 1-1 Namiki, Tsukuba 305-0044, Japan
8. Southern University of Science and Technology, Shenzhen 518055, P.R. China
9. Department of Physics, Hong Kong University of Science and Technology, Hong Kong, China

*E-mail: dmitri.efetov@lmu.de

## Table of Contents

## A.- Methods

### A.1.- Device fabrication

The TBG van der Waals heterostructures are fabricated using a "laser-and-stack" technique. Graphene and hBN flakes are first exfoliated on a $Si^{++}/SiO_2$ (285 nm) substrate and later picked up using a polycarbonate (PC)/polydimethylsiloxane (PDMS) stamp. The graphene flake is initially cut with an infrared laser of wavelength 1064 nm using a WITec alpha300 R Raman Imaging Microscope. The PC/PDMS stamp was used to pick-up subsequently the top hBN, first and second graphene layers, and finally the bottom hBN. Before picking up the second graphene layer, the stage is rotated by an angle of 1.1–1.2°. All layers were picked up at a temperature of ~100 °C. The finalized stack is dropped on a $Si^{++}/SiO_2$ substrate by melting the PC at 180 °C.

To fabricate the Josephson junctions, we spin coat a PMMA resist at 6000 rpm and bake at 150 °C for 2 min. After the e-beam lithography patterning, the resist development is carried out in an IPA:DI water (7:3) mixture at room temperature. The stack is shaped into a rectangular mesa by reactive ion etching using a $CHF_3/O_2$ mixture (40/4 sccm). The one-dimensional superconducting contacts are made by repeating the same process as with the mesa etch, followed by dc sputtering of NbTi/NbTiN (5 nm/110 nm) and lift-off in NMP at 80 °C.

An example of a device is shown in a micrograph in Fig. S1a, where the van-der-Waals heterostructure of hBN/TBG/hBN is contacted with multiple NbTiN leads. The geometrical dimensions of a typical junction are shown in Fig. S1b, where $L$ is the length and $W$ the width. The length of the superconducting electrodes across the junction is $L_S$. The superconducting gap of our NbTiN leads $\Delta \sim 2.1$ meV is deduced from its $T_c$ by following the BCS formula $\Delta \sim 1.764 k_B T_c$ (see Fig. S1c) [1].

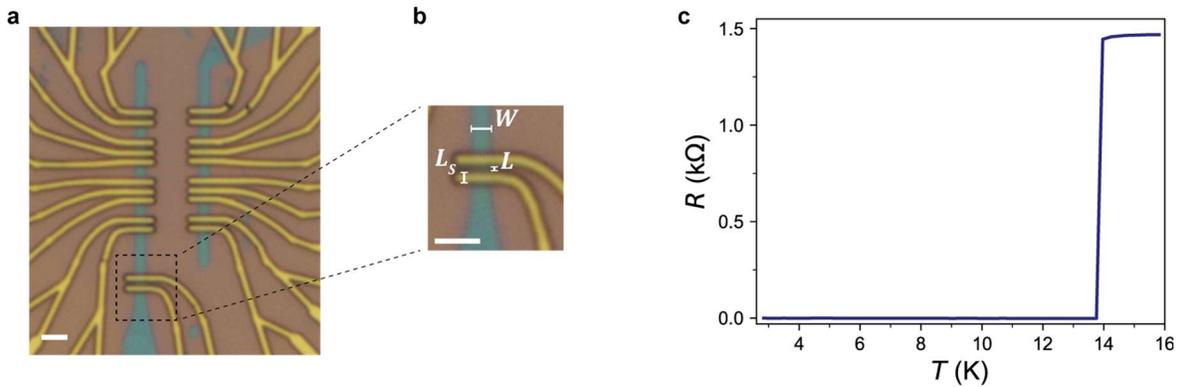

**Fig. S1. Device example and superconducting NbTiN. a,** Picture of a device, where multiple NbTiN (gold color leads) contacts are made to the van-der-Waals structure (blue color strips), which has already been etched in a rectangular mesa. **b,** Zoom-in of **a** showing the geometrical dimensions of a typical junction, with $L$ the length and $W$ the width. The length of the superconducting electrodes across the junction is $L_S$. White scale bars are 2 μm. **c,** Resistance as a function of temperature of a NbTiN film. The superconducting transition features a critical temperature $T_c \approx 13.8$ K.

## A.2.- Measurement techniques

The measurements were carried out in a dilution refrigerator (Bluefors SD250) with a base temperature of 35 mK. The d.c. measurements of the IV characteristics were performed by applying a d.c. bias current with a Keithley 2450 voltage source-meter in series with a 10 MΩ resistor. The resulting d.c. voltage was recorded in a multimeter. For the a.c. measurements we have used a standard low-frequency lock-in technique (Stanford Research SR860 amplifiers) with an excitation frequency f = 17.777 Hz. The differential resistance d$V$/d$I$ measurement was performed by biasing an a.c. excitation current varying from 0.5 nA to 2 nA, generated by the lock-in amplifier in combination with a 10 MΩ resistor. A d.c. bias current was applied through another 10 MΩ resistor before combining it with the a.c. excitation. The as-induced differential voltage and differential current were further measured at the same frequency with the standard lock-in technique. A Keithley 2400 source-meter was used to control the gate. All measured signals were filtered at the mixing chamber and still plate using commercially available low-pass RC and LC filters, respectively. After that, the signals were further filtered and amplified at room temperature by voltage-preamplifiers SR560 before entering the lock-in amplifiers or the multimeters. In measurements with a two-probe scheme, a finite known in-series resistance of 3447 Ω was subtracted from the data.

The magnetic field was applied by providing current to a superconducting magnet coil with a Keithley 2400. We achieve a field step size as small as a few microteslas. In order to minimize the amount of trapped flux in our superconducting magnet when measuring the interference patterns plots, we follow the next procedure. First, the magnetic field is gradually ramped from zero to the desired maximum positive value at a rate of ~ 0.2 mT/s. This is done by first stopping at a positive field of approximately a third of the target value, then sweeping in the opposite direction of the field down to a negative field of approximately two thirds of the target value, and finish by ramping up the field to the targeted maximum positive value. This results in an offset in the zero magnetic field with respect to the zero flux in the junctions between one and two data points in field, which is then subtracted from the raw data. At every point in field, in order to record the critical current of the junctions for both directions of the d.c. current, the d$V$/d$I$ is measured while sweeping the current from zero to the maximum positive value, then to zero again and finally swept to the maximum negative value.

## A.3.- Twist-angle extraction

The twist-angle is extracted from the high-field phase diagrams shown in Fig. S8. From the slope of the Landau levels in magnetic field vs. gate-voltage and their filling factor we can extract the carrier density. Then, by extracting the carrier density corresponding to a fully filled superlattice unit cell $n_s$, and applying the relation $n_s = 8\theta/\sqrt{3a^2}$, where $a$ = 0.246 nm is the graphene lattice constant; we extract a twist-angle $\theta$ with an uncertainty of 0.01°.

## B.- Additional information about devices D1-3

In this section we provide additional information about the junction devices D1-3 showed in this work, from their geometrical parameters, to their critical current, normal-state conductance and interference patterns across the different bands. Table S1 summarizes all the information about these.

| Device | $L$ (μm) | $W$ (μm) | $\theta$ (°) | $\Delta B$ (mT) | $\Delta B_{\text{phys}}$ (mT) | $I_c R_N$ (meV) | $l_{\text{mfp}}$ (nm) | $\xi_N$ (nm) |
|---|---|---|---|---|---|---|---|---|
| D1 | 0.20 | 1.00 | 0.94 | 3.6 | 3.4 | 0.16 | 36 | 51 |
| D2 | 0.20 | 1.50 | 1.00 | 2.5 | 2.3 | 0.11 | 35 | 45 |
| D3 | 0.15 | 1.50 | 1.24 | 2.8 | 2.5 | 0.12 | 23 | 33 |

**Table S1.** Summary of the three devices close to magic-angle measured in this work, D1-3. Shown parameters include length $L$ ($\pm$ 0.02 μm), width $W$ ($\pm$ 0.02 μm), twist-angle $\theta$ ($\pm$ 0.01°), measured field-periodicity of the critical current oscillations $\Delta B$ ($\pm$ 0.2 mT) and the expected periodicity from the physical device area and flux focusing effects $\Delta B_{\text{phys}}$ ($\pm$ 0.1 mT). Transport properties include the product of the critical current with the normal resistance $I_c R_N$, the mean-free-path $l_{\text{mfp}}$ and the superconducting coherence length in the weak-link $\xi_N$.

When a perpendicular magnetic field is applied to a Josephson junction featuring a homogenous supercurrent profile, the critical current follows an interference pattern where it vanishes every multiple integer multiple of flux quanta $\Phi_0 = h/2e$ that threads through the junction, such that $\Delta B_{\text{phys}} = \Phi_0/WL$. For our superconducting leads of NbTiN the London penetration depth is $\lambda_L \sim$ 400 nm, which is larger than the thickness of the film ($\sim$ 110 nm) and on the order of the length of the electrodes across the junction $L_S \sim$ 400 nm (see Fig. S1b) [2]. This implies that the magnetic field is partially penetrating the leads and thus changing the expected periodicity to $\Delta B_{\text{phys}} = \Phi_0/WL_{\text{eff}}$, where $L_{\text{eff}} = L + L_S$. This latter consideration matches well with the measured periodicity $\Delta B$ of the interference patterns of our devices (see Table S1).

By extracting the electron mean-free-path, $l_{\text{mfp}} = \hbar L\sqrt{\pi}/e^2 W R_N \sqrt{n}$, where $n$ is the electron carrier density, and comparing it to the length of our junctions $L$ (see Table S1), we conclude that all of them are in the diffusive regime since $l_{\text{mfp}} < L$. The superconducting coherence length inside the weak-link in the diffusive regime is expressed as $\xi_N = \sqrt{\hbar D/\Delta} = \sqrt{\hbar v_F l_{\text{mfp}}/2\Delta}$, where $v_F$ is the Fermi-velocity and $D$ the diffusion coefficient in two dimensions. In all samples we find that $\xi_N < L$, implying that they are in the long regime [3]. For a band structure such as that of TBG, where the Fermi velocity is not constant, we use the Einstein relation $R_N^{-1} = WL^{-1}De^2\text{DOS}$, thus getting $\xi_N = \sqrt{\hbar L/We^2\Delta\,R_N\,\text{DOS}}$. Thus, by using the DOS of the continuum model, we get the values of $\xi_N$ in the dispersive bands shown in Table S1. Since the mean free path is similar in the flat-bands, but the Fermi velocity is much lower, we conclude that when the Fermi level is tuned into the flat-bands the JJs fall even deeper in the long regime. This is also consistent with the $I_c R_N$ values being much smaller than $\Delta \sim$ 2.1 meV (see Table S1 and Fig. S5).

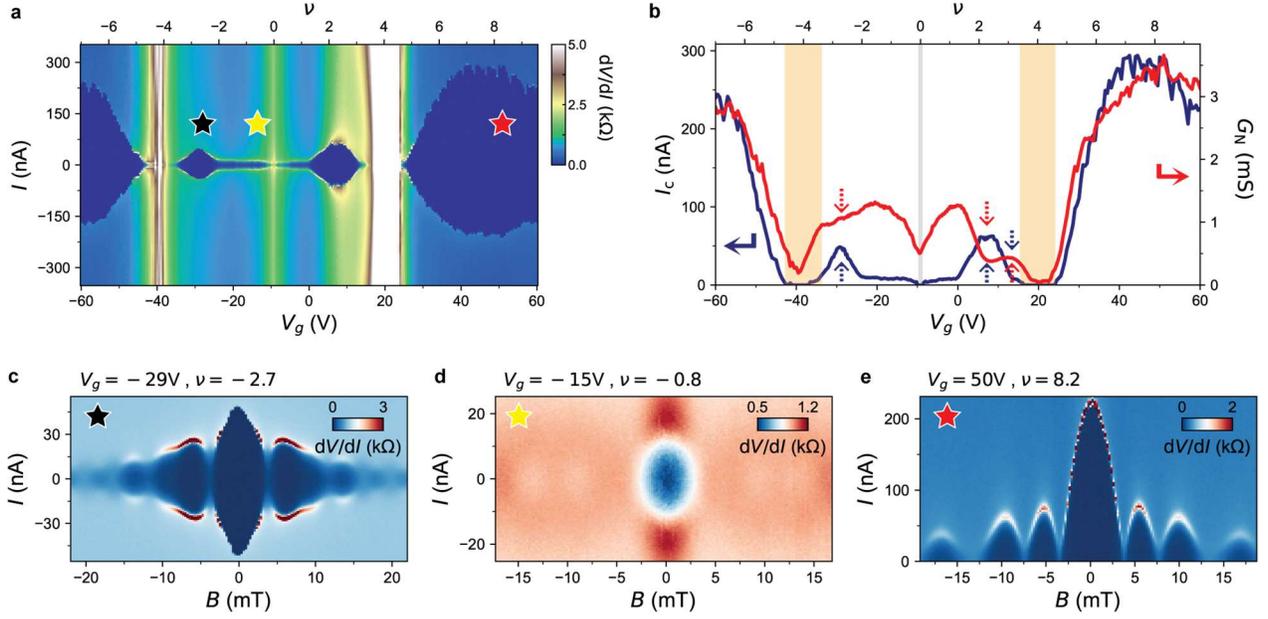

**Fig. S2. Differential resistance map and interference patterns of device D1. a,** d$V$/d$I$ as a function of d.c. current $I$, gate-voltage $V_g$ (bottom) and moiré filling factor $\nu$ (top). **b,** Critical current $I_c$ (blue, left axis) and normal state conductance $G_N$ (red, right axis) as a function of $V_g$ (bottom) and $\nu$ (top). **c-e,** Interference patterns at different dopings.

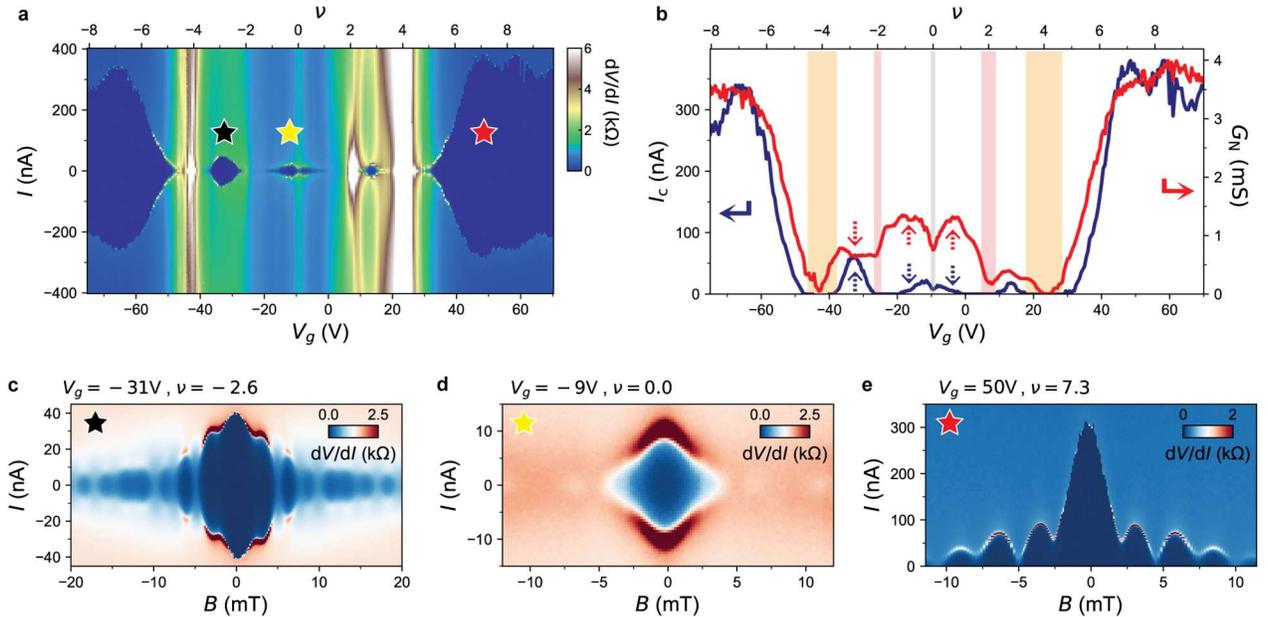

**Fig. S3. Differential resistance map and interference patterns of device D2. a,** d$V$/d$I$ as a function of d.c. current $I$, gate-voltage $V_g$ (bottom) and moiré filling factor $\nu$ (top). **b,** Critical current $I_c$ (blue, left axis) and normal state conductance $G_N$ (red, right axis) as a function of $V_g$ (bottom) and $\nu$ (top). **c-e,** Interference patterns at different dopings.

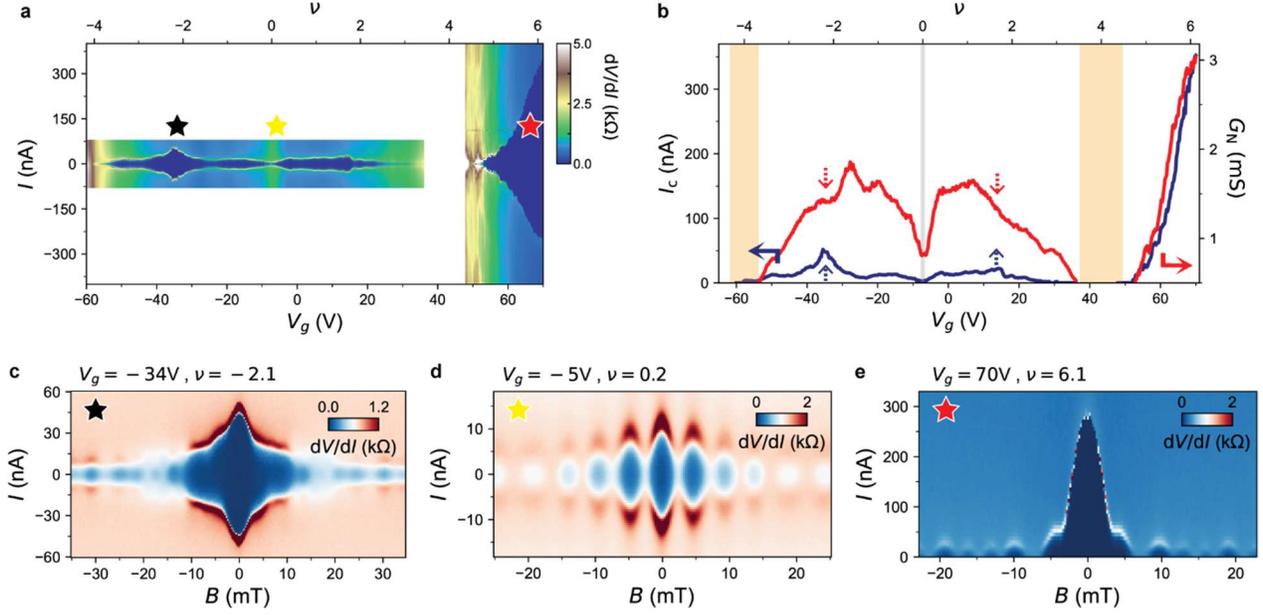

**Fig. S4. Differential resistance map and interference patterns of device D3. a,** d$V$/d$I$ as a function of d.c. current $I$, gate-voltage $V_g$ (bottom) and moiré filling factor $v$ (top). **b,** Critical current $I_c$ (blue, left axis) and normal state conductance $G_N$ (red, right axis) as a function of $V_g$ (bottom) and $v$ (top). **c-e,** Interference patterns at different dopings.

In Figs. S2a, S3a and S4a we provide the entirety of the d$V$/d$I$ maps of the junctions D1-3, covering both the flat and dispersive bands. In Figs. S2b, S3b and S4b, it can be seen that $I_c$ and $G_N$ are correlated between each other in the dispersive bands, but not in some regions of the flat-bands as indicated by dashed arrows.

The interference patterns at the dispersive bands show oscillations typical of a uniform, single slit JJ, as shown in Figs. S2e, S3e and S4e. At the flat-bands, near CNP, the oscillations are barely visible (Figs. S2d, S3d and S4d). We attribute this to their very low $I_c$ at zero field (~ 10 nA), which results in even lower values during the following oscillations in field (< 2 nA). These values of $I_c$ correspond to a Josephson energy of $E_J = \hbar I_c/2e$ ~ 4.1 µeV that is comparable to the thermal fluctuations $k_B T$ ~ 3.0 µeV at the base temperature of our fridge. Therefore, such low values of $I_c$ cannot be resolved in our experimental setup (the value of $k_B T$ represents a lower limit, since it would be higher if we considered the electronic temperature instead of the bath temperature). Furthermore, the thermal fluctuations can also cause phase diffusion [4], increasing the junction resistance at zero d.c. current to finite values, as observed. In the case of fillings $|v| > 2$ the observed interference patterns show asymmetric oscillations in all samples (Figs. S2c, S3c and S4c). How these latter evolve within the dome can be further checked in Supplementary section I.

As shown in Fig. 2 of the main text, the $I_c R_N$ product in the flat-bands can reach the same value as in the dispersive bands. Fig. S5 shows the reproducibility in D1. For D3, $I_c R_N$ only reaches 0.04 meV as in other fillings of D1 and D2. In Fig. S6 we also show two different methods of extracting $R_N$ (and therefore $G_N = R_N^{-1}$) that follow the same trends vs. $v$, and quantitatively both show violations of the correlation between $I_c$ and $G_N$.

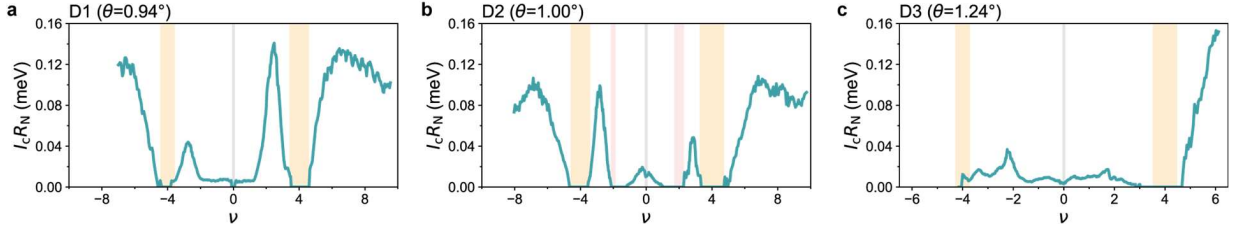

**Fig. S5.** $I_cR_N$ **product for devices D1-3. a-c,** Product of the critical current and normal-state resistance, $I_cR_N$, as a function of the moiré filling factor $v$, for devices D1, D2 and D3; respectively.

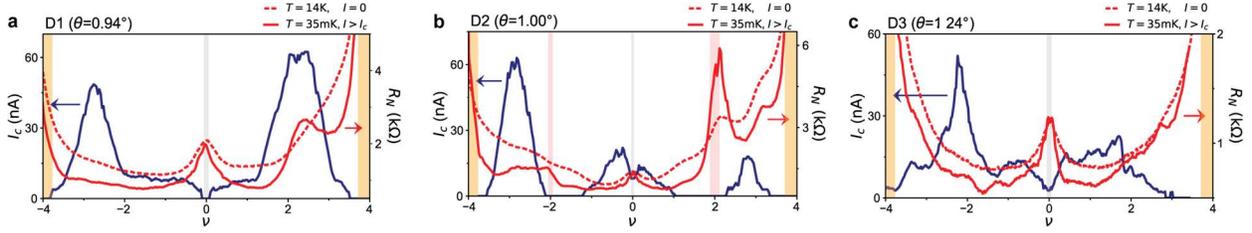

**Fig. S6. Normal-state resistance extraction. a-c,** $I_c$ (blue, left axis) and $R_N$ (red, right axis) vs. $v$ for devices D1-3. $R_N$ can be extracted by measuring resistance at 14 K (dashed lines) when NbTiN is no longer SC. Because the states at 14 K hardly correspond to the ground states of TBG at 35 mK, we also extract $R_N$ by measuring resistance at a finite d.c. current $I$ greater than $I_c$ (solid lines). In the latter case, the insulating states at half-filling are still present in D2. Both extractions show a qualitative agreement between each other and do not alter the conclusions of this work.

### C.- Transport characterization of devices D1-3

This section includes additional transport measurements of devices D1-3. These are resistance vs. temperature (Fig. S7) and Landau Fan diagrams (Fig. S8). The presence of correlated insulators at half-filling in D2 (Fig. S7b), and of correlated Chern insulators at high magnetic fields for all devices D1-3 (Fig. S8), shows that they are close to the magic angle. The quantization of these latter and their sequence $(C, v) = (4 - v, v)$, where $C$ is the Chern number, is visible in Fig. S8b,d,f. This is consistent the expected two-terminal resistance at sufficient high magnetic fields in the quantum Hall regime [5,6].

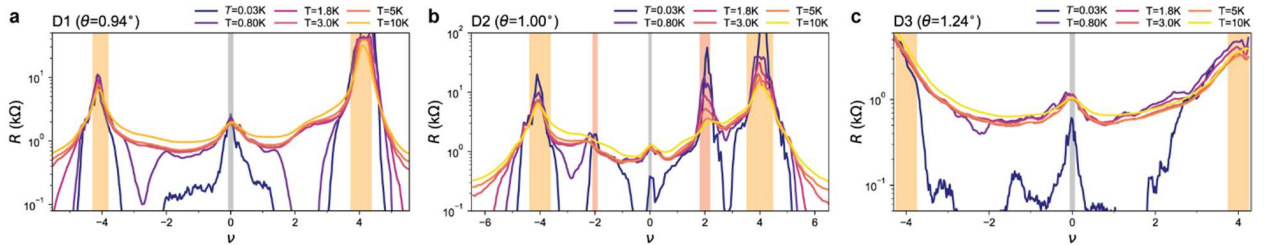

**Fig. S7. Resistance vs. temperature. a-c,** $R$ vs. $v$ for different temperatures $T$. In the dispersive bands all devices show a critical temperature $\sim 3$ K, and in general of $\sim 1$ K in the flat-bands.

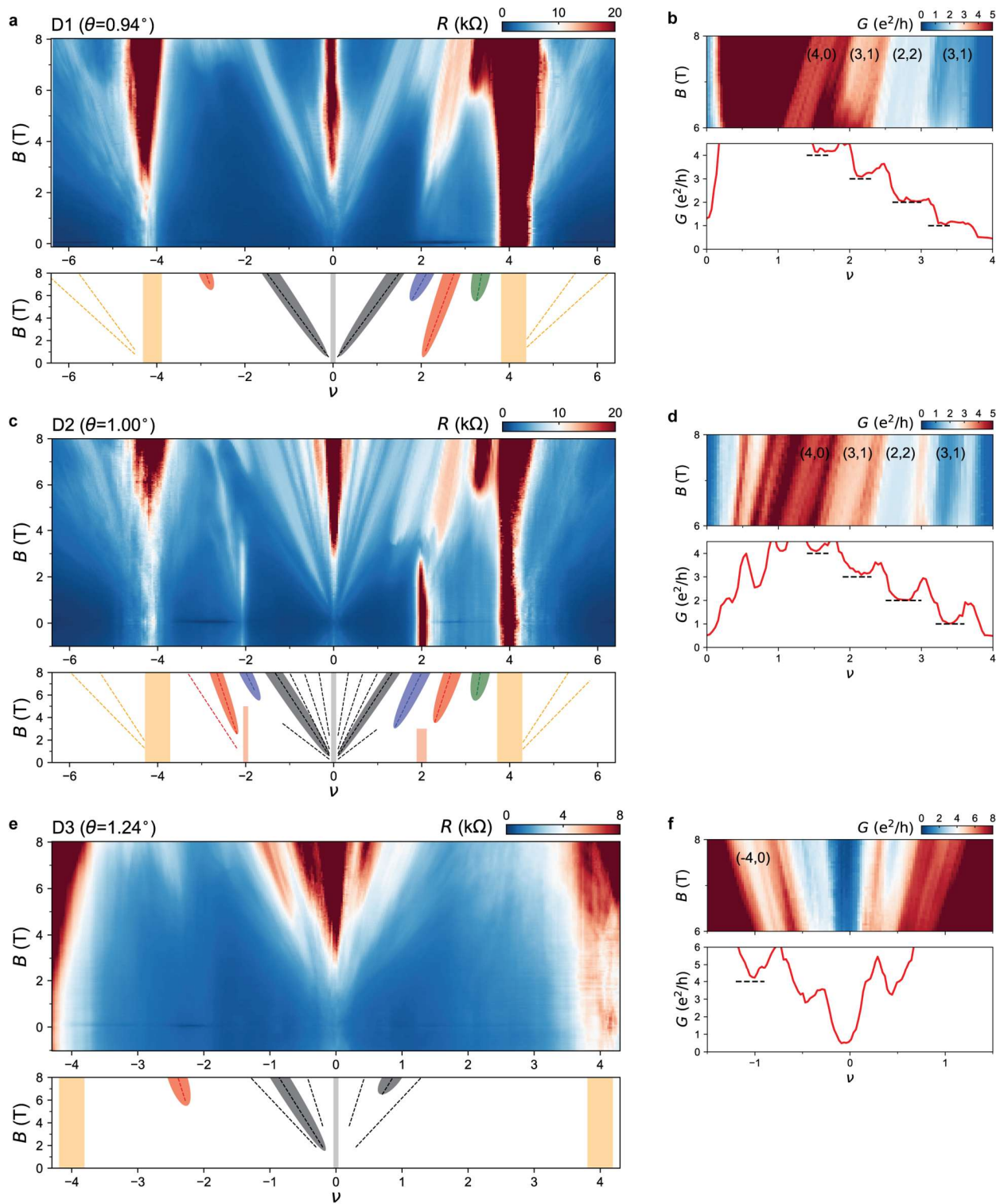

**Fig. S8. Landau fan of devices D1-3. a,** Resistance $R$ vs. moiré filling $v$ and magnetic field $B$ (top panel) along with the extracted Landau Levels in dashed lines (bottom panel), for D1. The shaded regions correspond to the observed Chern insulators. **b,** Top panel show a zoom-in of **a** at high field, in units of quantum of conductance. The bottom panel shows a line-cut at 8 T, with the quantization of the Chern insulators indicated by dashed horizontal lines and their sequence $(C, v)$ in the top panel. **c-d,** Analogous for device D2. **e-f,** Analogous for device D3.

## D.- Conventional features in TBG JJs away from magic angle (E1-3) & graphene JJs (G1)

This section shows data from additional devices not shown in the main text, such as their critical current, normal-state conductance and interference patterns across the different bands. These samples comprise TBG JJs further from the magic angle (E1-3) and monolayer graphene JJs (G1).

Table S2 summarizes all their information. Devices E1-3 and G1 follow the same architecture as D1-3, so their critical current periodicity with magnetic field $\Delta B$ also matches the prediction $\Delta B_{\text{phys}}$ when flux-focusing effects are accounted for and $L_S \sim 400$ nm.

Following the same methods described in Section B, we calculate the mean-free-path $l_{\text{mfp}} = \hbar L \sqrt{\pi}/e^2 W R_N \sqrt{n}$ and the coherence length $\xi_N = \sqrt{\hbar D/\Delta}$. In order to estimate $\xi_N$, for devices E1-3 we have used the DOS from the band structure of the continuum model, where in all cases we found that the junctions are in the long-diffusive regime. For G1 we have used the single layer graphene Dirac dispersion with $v_F \sim 10^6$ m/s., where $\xi_N$ is about twice as large compared to the TBG devices in Table S1 and Table S2. The $l_{\text{mfp}}$ is also about an order of magnitude larger, so G1 could correspond to an intermediate regime between short and long junctions.

| Device | $L$ (μm) | $W$ (μm) | $\theta$ (°) | $\Delta B$ (mT) | $\Delta B_{\text{phys}}$ (mT) | $I_c R_N$ (meV) | $l_{\text{mfp}}$ (nm) | $\xi_N$ (nm) |
|---|---|---|---|---|---|---|---|---|
| E1 | 0.15 | 1.50 | 0.45 | 2.5 | 2.5 | 0.10 | 20 | 98 |
| E2 | 0.25 | 1.50 | 0.51 | 2.1 | 2.1 | 0.05 | 36 | 101 |
| E3 | 0.25 | 1.50 | 0.72 | 2.1 | 2.1 | 0.05 | 31 | 89 |
| G1 | 0.45 | 4.50 | - | 0.5 | 0.54 | 0.10 | 196 | 155 |

**Table S2.** Summary of other devices measured in this work (E are the extra junctions further from magic-angle and G are the monolayer graphene junctions). Shown parameters include length $L$ ($\pm$ 0.02 μm), width $W$ ($\pm$ 0.02 μm), twist-angle $\theta$ ($\pm$ 0.01°), measured field-periodicity of the critical current oscillations $\Delta B$ ($\pm$ 0.2 mT) and the expected periodicity from the physical device area and flux focusing effects $\Delta B_{\text{phys}}$ ($\pm$ 0.1 mT). Transport properties include the product of the critical current with the normal resistance $I_c R_N$, the mean-free-path $l_{\text{mfp}}$ and the superconducting coherence length in the weak-link $\xi_N$.

In Fig. S9 we show the measurements of junctions E1-3. From the differential resistance maps Fig. S9b,e,h and their extraction of $I_c$ shown in Fig. S9a,d,g; we can see that the critical current follows a complicated trend vs. $v$. Nevertheless, $I_c$ does increase (decrease) when $G_N$ increases (decreases) and vice-versa, as expected for conventional Josephson junctions. We have also measured interference patterns at two different fillings for each device (see Fig. S9c,f,i), which follow the typical exponential decay of single-slit uniform junctions. They are also completely symmetric in current and field and no asymmetry due to a Josephson diode effect is recorded anywhere, in contrast with devices D1-3 that are closer to the magic-angle. More interference patterns at other fillings, not shown in here, have been measured with the same resulting symmetric patterns.

As for the single layer graphene JJ, G1, we find analogous results, both in the conventional correlation between $I_c$ and $G_N$ and in their symmetric interference patterns. This can be seen in Fig. S10.

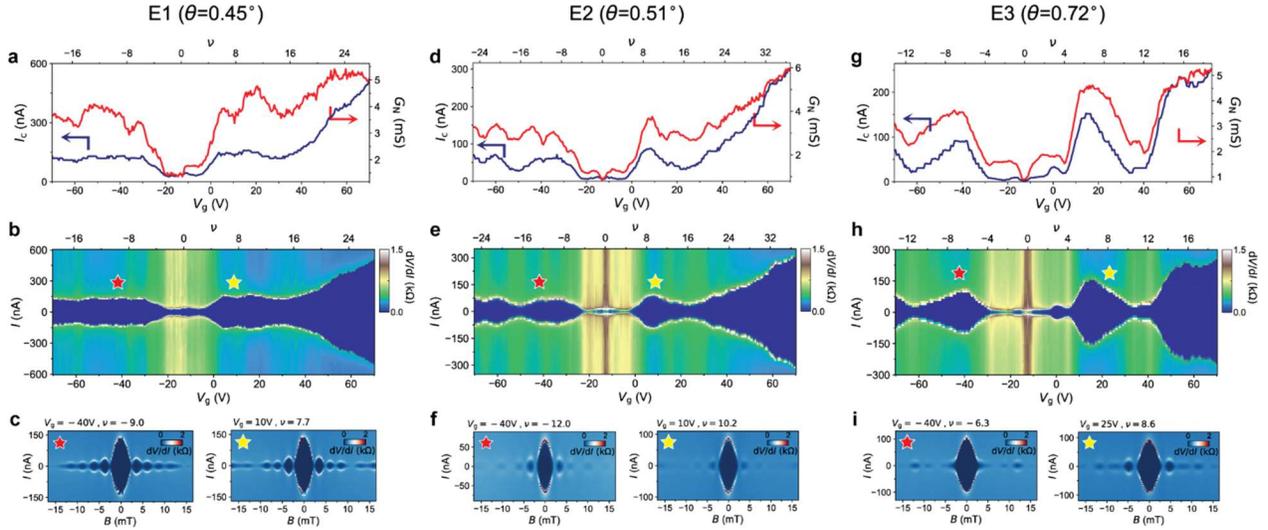

**Fig. S9. Additional TBG JJ devices E1-3 away from the magic angle. a,** Critical current $I_c$ (blue, left axis) and normal state conductance $G_N$ (red, right axis) as a function of gate-voltage $V_g$ (bottom) and filling $v$ (top) for device E1. **b,** Differential resistance $dV/dI$ as a function of d.c. current $I$, $V_g$ (bottom) and $v$ (top) for device E1. **c,** Interference patterns at two different fillings for device E1. **d-f,** Analogous for device E2. **g-i,** Analogous for device E3. All JJs show conventional behavior, both in their $I_c$ vs. $G_N$ relations and in their symmetric interference patterns.

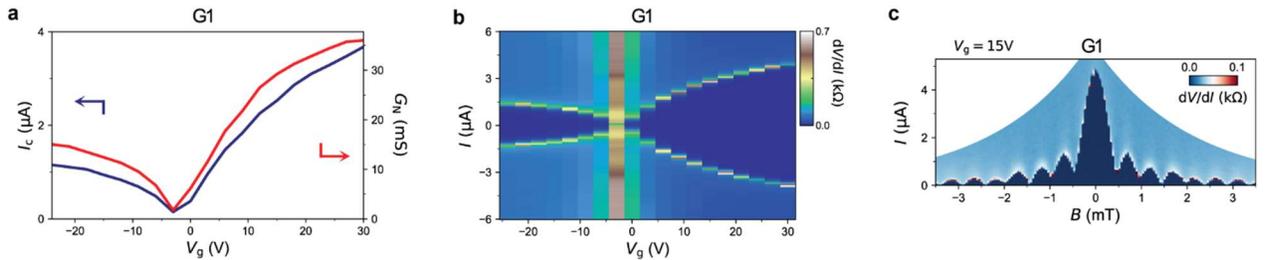

**Fig. S10. Monolayer graphene junction G1. a,** Critical current $I_c$ (blue, left axis) and normal state conductance $G_N$ (red, right axis) as a function of gate-voltage $V_g$. **b,** Differential resistance $dV/dI$ map vs. $V_g$ and vs. d.c. current $I$, measured at 300 mK. **c,** Interference pattern measured at 1 K. The graphene JJ shows conventional behavior both in its $I_c$ vs. $G_N$ relations and in their symmetric, uniform Fraunhofer interference patterns.

## E.- Theory of the interaction-induced part of the critical current

To illustrate how the interaction-induced part of critical current emerges, we consider a simple Ginzburg-Landau theory description at the flat-band limit as an example. The Ginzburg–Landau theory is found by the small $\Delta$ expansion of the free energy

$$F = \sum_i \frac{\Delta_i^2}{U} - T \sum_{\omega_n} \log \det[i\omega_n + \mu - H(\Delta)], \quad H(\Delta) = \begin{pmatrix} H_0 & \Delta \\ \Delta^* & -H_0 \end{pmatrix}$$

of an electron system with a mean field $\Delta_i$ describing the attractive interaction in each unit cell $i$ (uniform pairing on orbitals assumed), with attractive coupling $U > 0$. Here, $\omega_n = 2\pi T(n + 1/2)$ are Matsubara frequencies (for further details of the calculations in this section see Ref. [7]). Concentrating on the contribution from a single isolated exact flat-band, with Bloch Hamiltonian $H_0(k) = \epsilon_0 |k0\rangle\langle k0| + \cdots$, and at half-filling $\mu = \epsilon_0$, the Ginzburg-Landau expansion along the lines of Ref. [8,9] results to the free-energy density

$$F = \frac{\xi_g^2}{4TA_m} |\nabla \Delta|^2 + \frac{T - T_c}{4TT_c A_m} |\Delta|^2 + \frac{1}{96T^3 A_m} |\Delta|^4$$

Here mean-field $T_c = U/4k_B$ was identified, $A_m$ is the unit cell area, and $\xi_g^2$ is the Brillouin zone average of the (minimal) quantum metric. The SNS problem of minimizing $F$ at $T > T_c$ while keeping $\Delta(x < 0, y) = \Delta_S e^{-i\varphi/2}$, $\Delta(x > L, y) = \Delta_S e^{i\varphi/2}$ fixed can be analytically solved if the quartic $|\Delta|^4$ term is neglected. For a length $L$ large compared to $L_g = \xi_g \sqrt{U/(4k_B T - U)}$, the result is $F = \hbar I_c \cos(\varphi)/2e + \text{const.}$ and the critical current is

$$I_c^{\text{int}} = \frac{8e}{\hbar} \frac{W L_g}{A_m} \frac{\Delta_S^2}{U} \left(1 - \frac{U}{4k_B T}\right) e^{-L/L_g},$$

where $W$ is the system width in the y-direction [7]. The $|\Delta|^4$ term generally suppresses $I_c$ unless $T \gg U$. The value $\Delta_S$ of the mean field induced by the contact to superconducting lead can be smaller than the energy gap of the lead. The mean field describes the interactions, and generally is not continuous across material interfaces on the scale described by the Ginzburg–Landau theory. Note that $I_c^{\text{int}}$ does not dependent on the normal state conductance and is only the contribution to the $I_c$ that originates from the mean field $\Delta$, and it becomes zero when $U$ vanishes. The total $I_c$ also contains a component due to transport of electron pairs traversing the SNS junction without interactions; this is the component dependent on $G_N$ that is typically considered and is proportional to the Fermi velocity. However, in a quasi-flat-band it becomes small and therefore it is of interest to consider also $I_c^{\text{int}}$. In the simplest approximation, the two components simply sum together.

Quantitatively, $I_c^{\text{int}}$ approaches the experimental $I_c \sim 50$ nA by setting $U \sim 10$ µeV, $T \sim 100$ mK and the square root of the averaged Brillouin-zone integrated minimal quantum metric, $\xi_g$, to 40 nm, consistent with [8,9]. These results serve as an order of magnitude check that the typical energy, temperature and length scales of the experimental system can provide critical currents of the order of magnitude observed in the experiments, within the simple Ginzburg-Landau theory. A thorough analysis of the possibility of interaction-induced critical current would require a more microscopic theory of TBG interactions. The most important message of our Ginzburg-Landau theory approach is to show a possibility of critical current that does not depend on $G_N$.

# F.- Theory of contact-induced superconductivity into TBG

## F.1.- Derivation of the contact-induced pairing correlator

To be able to retain the full band structure and Bloch wavefunctions of the TBG continuum model [10], we focus on the contact-induced pairing correlator $\phi_\mathbf{R}$ in the middle of the junction. As we expect a monotonic relation between $\phi_\mathbf{R}$ and the critical current, the main features in the twist-angle and filling dependence of $I_c$ are expected to be captured by $\phi_\mathbf{R}$.

To this end, let us start by considering a general electron system without interactions and, thus, without intrinsic superconductivity, in which the pairing term is induced by the tunneling to an s-wave superconductor placed in its proximity. Assuming a structureless contact-induced pairing, the Hamiltonian can be written as follows

$$H = \int d\mathbf{r} c^\dagger_{\sigma,\eta}(\mathbf{r})[\hat{h}_{\sigma,\eta} - \mu]c_{\sigma,\eta}(\mathbf{r}) + \int d\mathbf{r}[\Delta(\mathbf{r})c^\dagger_{\uparrow,+}(\mathbf{r}) \cdot c^\dagger_{\downarrow,-}(\mathbf{r}) + \text{H. c.}],$$

where $c_{\sigma,\eta}(\mathbf{r})$ is the fermionic operator which annihilates an electron with spin $\sigma = \uparrow, \downarrow$ and valley $\eta = +, -$ at point $\mathbf{r}$; $\hat{h}_{\sigma,\eta}$ is the single-particle Hamiltonian; $\mu$ is the chemical potential; $\Delta(\mathbf{r})$ is the contact-induced pairing amplitude; and $\cdot$ corresponds to the inner product (diagonal summation over internal degrees of freedom, e.g., sublattice and layer in the case of TBG). As said, in our analysis, we are primarily interested in the pairing correlator given by

$$\phi(\mathbf{r}) = \langle c_{\downarrow,-}(\mathbf{r}) \cdot c_{\uparrow,+}(\mathbf{r}) \rangle.$$

In the considered setup, this quantity can be also thought of as a response function to the external superconducting field $\Delta(\mathbf{r})$.

Let us move to the Bloch-state basis,

$$c_{\sigma,\eta}(\mathbf{r}) \to \sum_{\mathbf{k}\in \text{B.z.},n} e^{i\mathbf{k}\mathbf{r}} u_{\mathbf{k},\sigma,\eta,n}(\mathbf{r}) c_{\mathbf{k},\sigma,\eta,n},$$

where $\mathbf{k}$ is a wave-vector in the Brillouin zone (B.z.) and $n$ denotes the band index. The Bloch amplitude $u_{\mathbf{k},\sigma,\eta,n}(\mathbf{r})$ is a vector with components corresponding to the internal degrees of freedom – again, for the TBG, these are sublattice and layer. Keeping different bands in the summation over $n$, we can investigate the projection onto the active bands and study which of them play the crucial role in the formation of the superconductivity. In the band basis, the Hamiltonian reads

$$H = \sum_{\mathbf{k},\sigma,\eta,n} c^\dagger_{\mathbf{k},\sigma,\eta}[\epsilon_{k,\sigma,\eta} - \mu]c_{\mathbf{k},\sigma,\eta,n} +$$

$$\int d\mathbf{r}\Delta(\mathbf{r})\left[\sum_{\mathbf{k},\mathbf{k}'} e^{i(\mathbf{k}'-\mathbf{k})\mathbf{r}} \sum_{n,n'} u^\dagger_{\mathbf{k},\uparrow,+,n}(\mathbf{r}) \cdot u^\dagger_{-\mathbf{k}',\downarrow,-,n'}(\mathbf{r})c^\dagger_{\mathbf{k},\uparrow,+,n}c^\dagger_{-\mathbf{k}',\downarrow,-,n} + \text{H. c.}\right]$$

We now treat the pairing term as a perturbation and obtain straightforwardly the SC correlator linearly dependent on $\Delta(\mathbf{r})$ in the imaginary time path integral formalism,

$$\phi(\mathbf{r}) = \sum_{\mathbf{k},\mathbf{k}'} \sum_{n,n'} \sum_{\omega_m} G_0\left(i\omega_m, \xi_{\mathbf{k},+,n}\right) G_0\left(-i\omega_m, \xi_{-\mathbf{k}',-,n'}\right) \times$$
$$\int d\mathbf{r}' \Delta(\mathbf{r}') e^{i(\mathbf{k}-\mathbf{k}')(\mathbf{r}-\mathbf{r}')} u^{\dagger}_{-\mathbf{k}',-,n'}(\mathbf{r}) u_{\mathbf{k},+,n}(\mathbf{r}) u^{\dagger}_{\mathbf{k},+,n}(\mathbf{r}') u_{-\mathbf{k}',-,n'}(\mathbf{r}'),$$

where we have assumed that the system is spin degenerate and the pairing amplitude is static; $G_0(i\omega_m, \xi_{k,\eta}) = (i\omega_m - \xi_k)^{-1}$ with $\xi_{\mathbf{k},\eta,n} = \epsilon_{\mathbf{k},\eta,n} - \mu$ and the fermionic Matsubara frequency $\omega_m = \pi(2m+1)T$. Since we are interested in studying TBG, we apply the spinless time-reversal symmetry which requires $\xi_{\mathbf{k},\eta,n} = \xi_{-\mathbf{k},-\eta,n}$ and $u^*_{\mathbf{k},\eta,n} = u_{-\mathbf{k},-\eta,n}$; hence, the expression for the correlator can be rewritten within one valley (for later we fix the valley $\eta = +$ and make it a silent index, i.e., $\xi_{\mathbf{k},+,n} =: \xi_{\mathbf{k},n}$ and similarity for $u$). After summation over the Matsubara frequencies, we get

$$\phi(\mathbf{r}) = \sum_{\mathbf{k},\mathbf{k}'} \sum_{n,n'} \frac{n_F(-\xi_{\mathbf{k}',n'}) - n_F(\xi_{\mathbf{k},n})}{\xi_{\mathbf{k}',n'} + \xi_{\mathbf{k},n}} \int d\mathbf{r}' \Delta(\mathbf{r}') e^{i(\mathbf{k}-\mathbf{k}')(\mathbf{r}-\mathbf{r}')} u^{\dagger}_{\mathbf{k}',n'}(\mathbf{r}) u_{\mathbf{k},n}(\mathbf{r}) u^{\dagger}_{\mathbf{k},n}(\mathbf{r}') u_{\mathbf{k}',n'}(\mathbf{r}'),$$

with $n_F(\xi)$ being the Fermi-Dirac distribution function. We now assume that the induced pairing amplitude changes slowly within the unit cell, therefore, $\Delta(\mathbf{R} + \mathbf{a}) \approx \Delta(\mathbf{R})$, where $\mathbf{R}$ is the lattice vector and $\mathbf{a}$ lies inside the chosen unit cell (u.c.). Furthermore, since we are interested in the large-scale spatial behavior, we study the correlator averaged over the u.c., $\phi_{\mathbf{R}} = \int_{\text{u.c.}} d\mathbf{a} \, \phi(\mathbf{R} + \mathbf{a})$. Finally, making the symmetrical substitution $\mathbf{k} \to \mathbf{k} + \mathbf{q}/2$, $\mathbf{k}' \to \mathbf{k} - \mathbf{q}/2$, we arrive at the main form of the studied SC correlator,

$$\phi_{\mathbf{R}} = \sum_{\mathbf{q},\mathbf{k}} \sum_{\mathbf{R}'} e^{i\mathbf{q}(\mathbf{R}-\mathbf{R}')} \Delta(\mathbf{R}') \sum_{n,n'} \phi^{\text{disp}}_{\mathbf{k},\mathbf{q},n,n'} \phi^{\text{QG}}_{\mathbf{k},\mathbf{q},n,n'},$$

$$\phi^{\text{disp}}_{\mathbf{k},\mathbf{q},n,n'} = \frac{\tanh[\xi_{\mathbf{k}-\mathbf{q}/2,n'}/(2T)] + \tanh[\xi_{\mathbf{k}+\mathbf{q}/2,n}/(2T)]}{\xi_{\mathbf{k}-\mathbf{q}/2,n'} + \xi_{\mathbf{k}+\mathbf{q}/2,n}}, \quad \text{(S1)}$$

$$\phi^{\text{QG}}_{\mathbf{k},\mathbf{q},n,n'} = \left| \int_{\text{u.c.}} d\mathbf{a} \, e^{i\mathbf{q}\mathbf{a}} u^{\dagger}_{\mathbf{k}-\mathbf{q}/2,n'}(\mathbf{a}) u_{\mathbf{k}+\mathbf{q}/2,n}(\mathbf{a}) \right|^2.$$

Here, $\phi^{\text{disp}}_{\mathbf{k},\mathbf{q},n,n'}$ represents the kinetic contribution to the induced SC. The second term $\phi^{\text{QG}}_{\mathbf{k},\mathbf{q},n,n'}$ involves the overlap between Bloch states at different momenta and is, thus, determined by the quantum geometry of the bands. Therefore, the correlator $\phi_{\mathbf{R}}$ can be expressed as the Fourier transform of the product of this contributions.

Seeking a deeper theoretical understanding of the importance of the quantum geometry and multiband effects, we define several limits. First, the "flat-band limit" is obtained by keeping in the summation over $n, n'$ only the two lowest bands of TBG and then substituting $\xi_{\mathbf{k},n} \to \alpha \xi_{\mathbf{k},n}$, $\alpha \to 0$. In this limit, we obtain

$$\phi_{\mathbf{R}} \to \frac{1}{2T} \sum_{\mathbf{q}} \sum_{\mathbf{R}'} e^{i\mathbf{q}(\mathbf{R}-\mathbf{R}')} \Delta(\mathbf{R}') \sum_{n,n'} \sum_{\mathbf{k}} \left| \int_{\text{u.c.}} d\mathbf{a} \, e^{i\mathbf{q}\mathbf{a}} u^{\dagger}_{\mathbf{k}-\mathbf{q}/2,n'}(\mathbf{a}) u_{\mathbf{k}+\mathbf{q}/2,n}(\mathbf{a}) \right|^2,$$

If $\Delta(\mathbf{R}')$ only has local support, then for large $\mathbf{R}$: $|\mathbf{R} - \mathbf{R}'| \gg \sqrt{V_{u.c.}}$, terms with small $\mathbf{q}$ are expected to have the dominant contribution. The small-$\mathbf{q}$ expansion allows to clearly see the connection to the quantum geometric tensor,

$$\phi_{\mathbf{R}} \approx \frac{1}{2T}\sum_{\mathbf{q}}\sum_{\mathbf{R}'} e^{i\mathbf{q}(\mathbf{R}-\mathbf{R}')} \Delta(\mathbf{R}') \sum_{n,n'}\sum_{\mathbf{k}} \left[\delta_{n,n'} - g_{\mu\nu,nn'}(\mathbf{k})q^{\mu}q^{\nu} - A_{\mu\nu}(\mathbf{k})q^{\mu}q^{\nu} + o(|\mathbf{q}|^4)\right],$$

where $g(\mathbf{k})$ is the multiband quantum metric,

$$\begin{aligned} g_{\mu\nu,nn'}(\mathbf{k}) &= \text{Re}[\delta_{n,n'}\langle\partial_{\mu}u_{\mathbf{k},n}|\partial_{\nu}u_{\mathbf{k},n}\rangle - \beta_{\mu,nn'}\beta_{\nu,n'n}], \quad \beta_{\mu,nn'} = i\langle u_{\mathbf{k},n}|\partial_{\mu}u_{\mathbf{k},n'}\rangle, \\ A_{\mu\nu,nn}(\mathbf{k}) &= \delta_{n,n'}\left[\delta_{\mu\nu}\langle u_{\mathbf{k},n}|a_{\mu}^2|u_{\mathbf{k},n}\rangle + 2\text{Im}\langle u_{\mathbf{k},n}|a_{\nu}|\partial_{\mu}u_{\mathbf{k},n}\rangle\right] - \\ &\quad 2\text{Re}[\langle u_{\mathbf{k},n'}|a_{\mu}|u_{\mathbf{k},n}\rangle\beta_{n'n,\nu}] + \langle u_{\mathbf{k},n'}|a_{\mu}|u_{\mathbf{k},n}\rangle\langle u_{\mathbf{k},n}|a_{\nu}|u_{\mathbf{k},n'}\rangle. \end{aligned}$$

Thus, we reveal the connection between the quantum metric and the formation of the contact-induced superconductivity. It is worth noting that, in the flat-band limit of Eq. (S1) at finite $T$, the filling factor of the flat-bands loses its meaning; therefore, this limit should be considered as a way to investigate the combined contribution of the geometrical structure of the Bloch states in the fully filled flat-bands.

Another paradigmatic limit is the "atomic limit", where the Bloch states are completely momentum-independent, exactly like in an atomic insulator. In this limit, we get $\phi_{\mathbf{k},\mathbf{q},n,n'}^{QG} = \delta_{n,n'}$. This limit not only removes the geometry of the states, allowing to focus on dispersion-driven effects, but also suppresses the interband superconducting contributions to the correlator completely.

From these expressions we can get the different contributions discussed in the main text and shown in Fig. S11. As a first approach, in the atomic limit, we only consider pairs generated at the electron flat-band (Fig. S11a), where the Fermi level $E_F$ lies. This means that $\phi_R^{S.B.} \sim \sum_{\mathbf{q},\mathbf{k}} \phi_{\mathbf{k},\mathbf{q},n,n}^{disp}$, were we do not sum over the index $n$. We find that this case follows well the density of states of the flat-bands for all three cases but differs significantly from the experimental $I_c$, as seen in Fig. S11e-g. When including pairs from other bands not located at $E_F$, their contribution to $\phi_R^{M.B.} \sim \sum_{\mathbf{q},\mathbf{k}}\sum_n \phi_{\mathbf{k},\mathbf{q},n,n}^{disp}$ diminishes the further these bands are from $E_F$ and the larger their dispersion $E_{k,n}$ is. Note that here we sum over the index $n$. Such multiband effect is then expected to be particularly significant in the flat-bands due to their close proximity in energy and low dispersion (Fig. S11c), as opposed to the high dispersive bands. Indeed, considering this mechanism we find that $\phi_R$ shifts towards higher fillings and adopts dome-like shapes centered around filling fractions $|\nu|$ that match our observations (Fig. S11k-m). The change is most notable in D1 and D2 than in D3, consistent with their lower bandwidths.

By now departing from the atomic limit we can now quantify other effects than dispersion, such as the non-trivial quantum metric of the bands (Fig. S11b). Considering the dispersion of the band where the $E_F$ lies and considering the quantum metric, $\phi_R^{Q.G.} \sim \sum_{\mathbf{q},\mathbf{k}} \phi_{\mathbf{k},\mathbf{q},n,n}^{disp} \phi_{\mathbf{k},\mathbf{q},n,n}^{QG}$. In Fig. S11h-j we see that the departure from the DOS—and thus from single-band process in the atomic limit— is most notable when the bandwidth of the flatband is smaller. This also approximates better to the experimental $I_c$, with the maximum of the correlator shifting towards higher fillings.

We finally note if both quantum geometry and multiband effects are, an interband effect also weighs in, where $n \neq n'$ and the Andreev pairs are formed by two electrons coming from two different bands (Fig. S11d). This results in $\phi_R \sim \sum_{\mathbf{q},\mathbf{k}} \sum_{n,n'} \phi^{\text{disp}}_{\mathbf{k},\mathbf{q},n,n'} \phi^{\text{QG}}_{\mathbf{k},\mathbf{q},n,n}$, giving broader single domes around $|\nu| \gtrsim 2$, as shown in Fig. S11n-p.

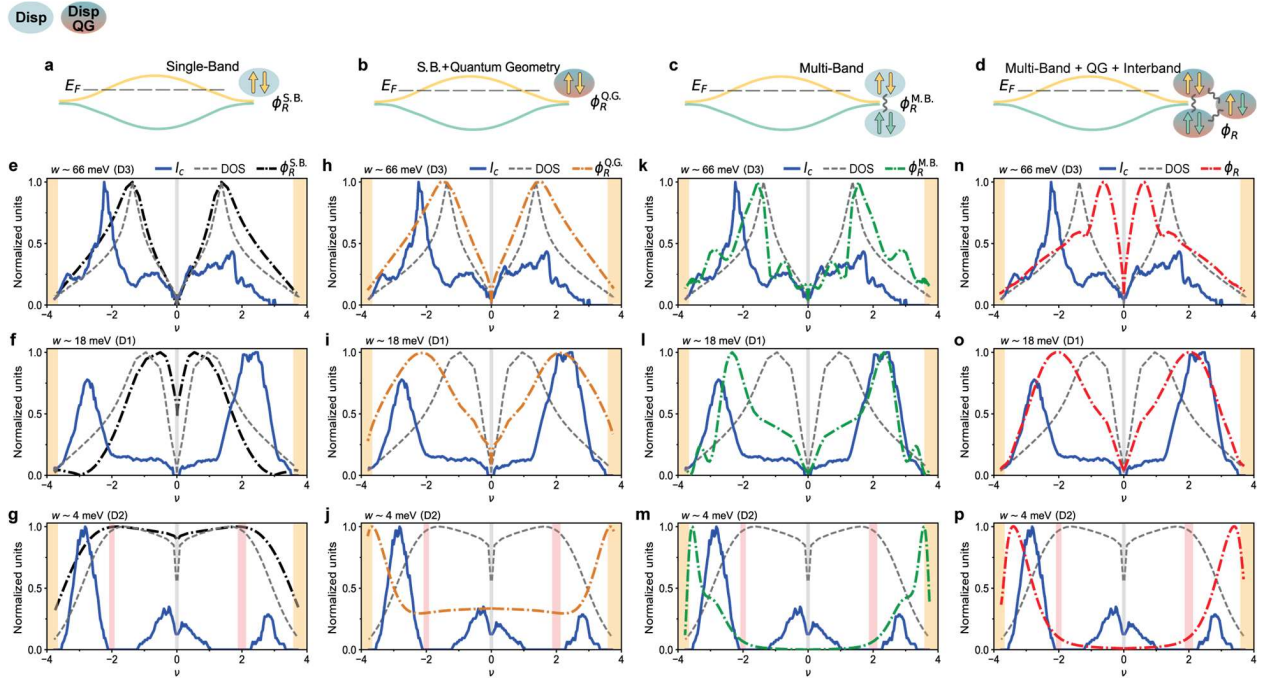

**Fig. S11. Different contributions to the induced superconductivity in the flat-bands. a-d,** Sketches illustrating the different contributions to the induced superconductivity. For **a** only the single-band process is considered. In **b**, the quantum geometry of the band is added. In **c**, multiband effects are taken into account. In **d**, all previous cases are involved, but on top of that, the interband processes are inevitably added, as explained above in the text. The color of the spin of the electrons forming the Andreev pair indicates which bands it is coming from. The color of the circles represents whether only the dispersion, or both the dispersion and the quantum geometry are being considered. **e-g,** Respectively for devices D3, D1 and D2, measured critical current $I_c$ (blue-solid line), along with the density of states (gray-dashed line) and the computed superconducting correlator for the single-band where the Fermi lies only $\phi_R^{\text{S.B.}}$ (black dash-dotted line); all as a function of the filling factor $\nu$ and in normalized units. **h-j,** Analogous to **e,f,g** but now the quantum geometry has been added to the correlator ($\phi_R^{\text{Q.G.}}$ in ochre dash-dotted line). **k-m,** Analogous to **e,f,g** but now the interference with more bands are accounted, i.e. multiband processes ($\phi_R^{\text{M.B.}}$ in green dash-dotted line). **n-p,** Analogous to **k,l,m** but by further considering the quantum geometry of these flat-bands, the interband pairing is also non-zero, giving the resulting correlator in red dash-dotted line.

We have shown the importance of the unconventional quantum geometric and multiband effects when describing the SC proximity effect in the flat-bands. In the case of the dispersive bands, the primary contributions do not include unconventional terms. Since the dispersive bands have large characteristic group velocities, the contact-induced superconductivity is mainly determined by the electrons in the small (relative to the bandwidth) vicinity of the Fermi surface. We should clarify here that the multiband effect is also present in the dispersive bands when and only when the Fermi surface crosses multiple bands. However, in stark contrast to the flat-bands, here, the proximity effect is expected to be well described by the energetically local quantity at the Fermi surface. Motivated by the fact that the overlap between the Bloch functions from different bands is generally small ($\phi^{QG}_{\mathbf{k},\mathbf{q}=\mathbf{0},n,n'} \sim \delta_{n,n'}$) we can also neglect the interband pairing processes. To demonstrate the conventional nature of the dispersive bands' proximity effect, we show in Fig. S12 that the DOS of the continuum model captures well the qualitative behavior of the measured $I_c$.

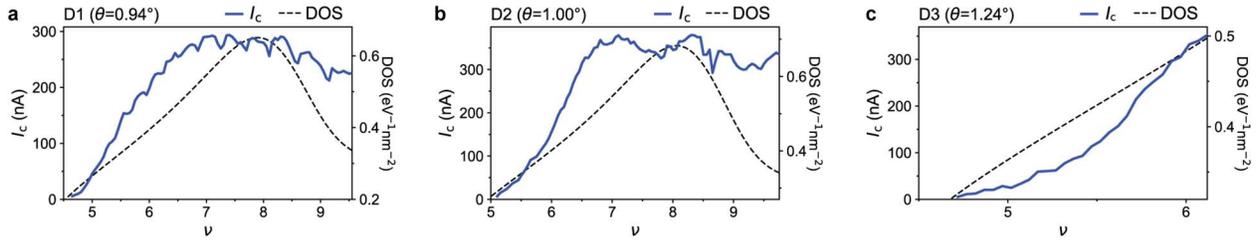

**Fig. S12. Proximity effect in the dispersive bands. a-c,** Extracted critical current $I_c$ (solid blue lines, left axis), along with the density of states DOS of the continuum model (black dashed lines, right axis) as a function of filling $\nu$ in the dispersive bands; respectively for devices D1, D2, D3.

### F.2.- Comments on the numerical calculations

In the numerical calculations presented in the main text, we used periodic boundary conditions, which is equivalent to working with the finite cluster of the Bloch states defined by the vectors $\mathbf{k} = n_1/N\mathbf{G}_1 + n_2/N\mathbf{G}_2$; here $n_{1,2} = -N/2, \ldots, N/2 - 1$, and $\mathbf{G}_{1,2}$ are basis vectors of the reciprocal lattice.

In order to obtain the bands structure and Bloch states of TBG, we used the continuum model [10] with parameters $t = 2.97$ eV (nearest neighbor hopping amplitude of the single layer graphene), $w_{AA} = 110$ meV, $w_{AB} = 80$ meV (inter-layer tunneling amplitudes), and the cutoff of the reciprocal grid lattice, commonly used in the plane wave expansions. The truncated Hamiltonian comprised 676 bands.

As mentioned, by continuity the critical current should be a monotonically increasing function of the SC correlator in the middle of the Josephson junction. Therefore, to make a comparison between the numerically obtained SC correlator and the critical current more universal and less dependent on the specific experimental geometry, we assumed that the induced pairing amplitude is localized inside one moiré unit cell, $\Delta(\mathbf{R}) = \Delta_0 \delta_{\mathbf{R},\mathbf{0}}$. Then for sufficiently large distances from the contact, the SC correlator is almost isotropic. Finally, we study the dependence of the correlator

$\phi_{\mathbf{R}}$ taken at the distance $|\mathbf{R}|$ corresponding to the middle of the junction studied in the experiment. Effects of boundary conditions were made negligible in our numerics by taking the system size to be $100 \times 100$ moiré unit cells, which is substantially larger than characteristic size of the devices used in the experiment.

### F.3.- Additional numerical results

As an extension of the numerical findings presented in the main text, Fig. S13 illustrates the dependence of the SC correlator within the flat-bands on both the twist-angle of the TBG and the filling factor. The left panels display $\phi$ calculated (Fig. S13a) within the atomic limit, and (Fig. S13b) when the full structure of the Bloch states is retained. In the summation over the bands in the equation (S1), we kept only flat-bands. It can be justified by the assumption that the superconductivity of the NbTiN lead has phonon-mediated origin [11], and therefore, the contact-induced pairing amplitude $\Delta(\mathbf{r})$ has a natural cutoff energy given by the Debye frequency ~50 meV.

Consistent with experimental observations, the SC correlator is suppressed near the magic angle (for the considered numerical parameters it is $\approx 1.01°$). To better understand the filling dependence of $\phi_{\mathbf{R}}$, in Fig. S13c-d we present the same results with different normalizations. Specifically, for each twist-angle, the correlator is normalized to its maximum value relative to the filling factor. We observe that the atomic limit qualitatively captures the critical current's dome-shaped behavior, including its peak's location. The incorporation of quantum geometry into the analysis results into a broadening of these dome-shaped regions.

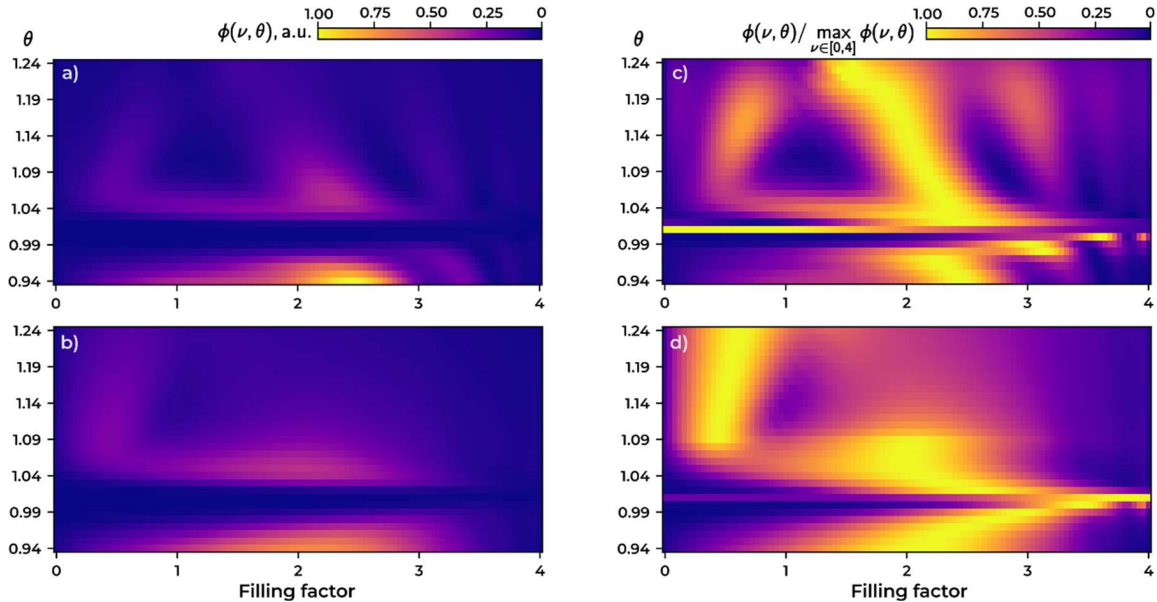

**Fig. S13. Contact-induced superconducting correlator vs. twist-angle and filling factor. a,** within the atomic limit. **b,** when the quantum geometry contribution is also taken into account. **c-d,** The same results with the row-normalization. The correlator is taken at $|\mathbf{R}| = 8$ moiré lattice vectors which corresponds to the center of the junctions used in the experiment.

## G.- Critical current extraction in the interference patterns

We define $I_c$ as the d.c. current bias value at which the $dV/dI$ is 30%–40% of the normal state resistance. When the magnetic field increases, the $dV/dI$ may not be exactly zero at zero d.c. current, which complicates the extraction of $I_c$.

Another approach can be taken for these cases. First, we locate the position of the nodes by extracting the local maxima of the $dV/dI$ vs. $B$ curve at zero current. We do this by calculating the intervals at which the second derivative of this curve, $d^2(dV/dI)/dB^2$, takes negatives values. We then set two possible conditions for setting the value of $I_c$: it is the $I$ value at which the $dV/dI$ is 30%–40% of the normal state resistance, or the $I$ value at which $d^2(dV/dI)/dB^2$ takes negatives values beyond a threshold; whichever is found first. The second condition over the second derivative helps finding the nodes, where $I_c = 0$, and the decay of the oscillations at low values of $I$. Fig. S14a-b shows how this method captures very well the oscillations with least resistance in the interference pattern (dark blue).

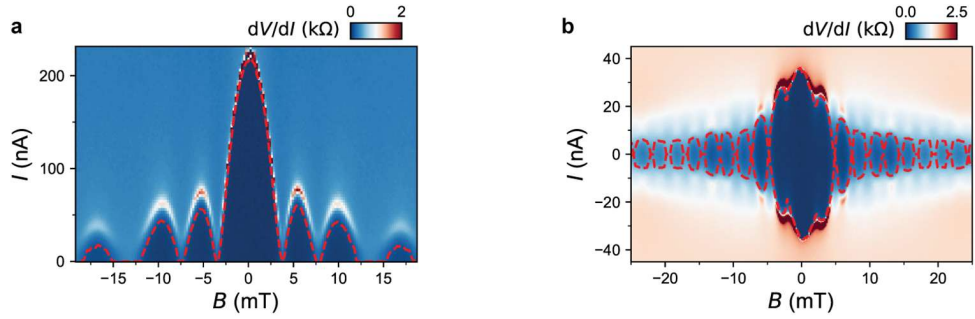

**Fig. S14. Critical current extraction. a,** Interference pattern of D1 at $V_g = 50$ V ($v = 8.2$) in the dispersive bands, along with the extracted $I_c$ in dashed red lines. **b,** Interference pattern of D2 at $V_g = -30$ V ($v = -2.5$), along with the extracted $I_c$ in dashed red lines. In both cases $I_c$ is taken as the $I$ value at which the resistance is 35% of the normal-state resistance, or the $I$ value at which $d^2(dV/dI)/dB^2$ takes negatives values.

## H.- Additional Josephson diode effect measurements for all samples

The reproducibility of the reversible JDE is presented on Fig. S15 and Fig. S16. The former shows the rectification measurements of device D2 at different fields. For D1 and D3, we have not performed rectification measurements, but we present the measurements of the d$V$/d$I$ curves for opposite $B$, where the two relations $I_c^+(B) \neq |I_c^-(B)|$ and $I_c^+(B) = |I_c^-(-B)|$ are clearly observed.

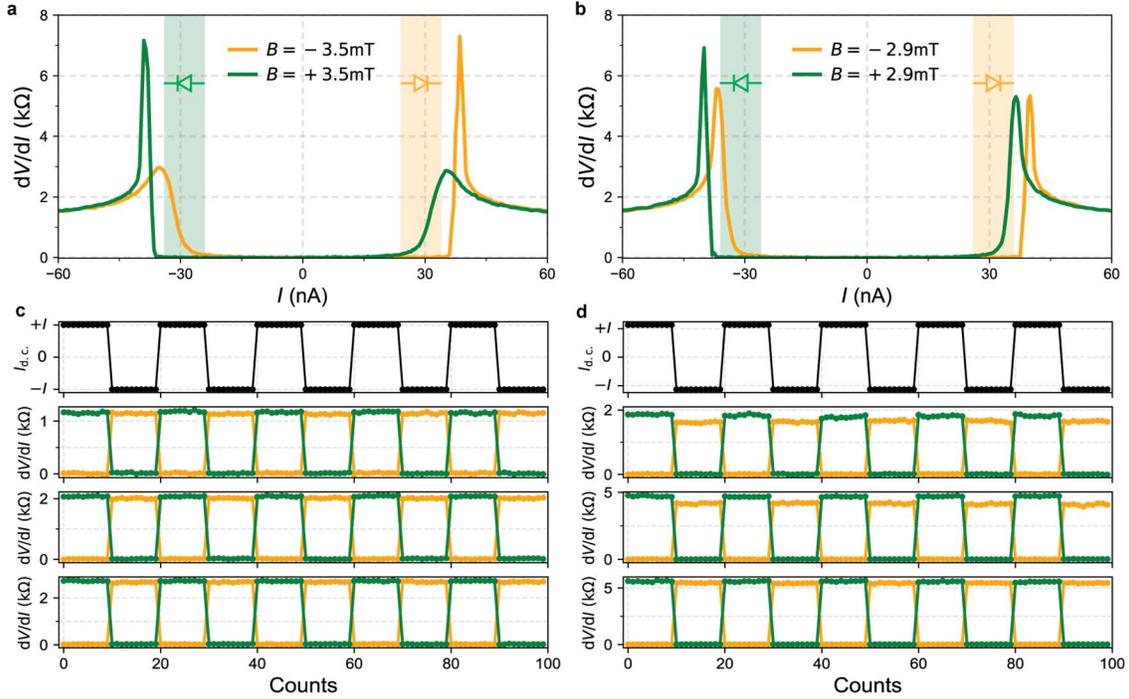

**Fig. S15. Additional Josephson diode rectification measurements. a,c,** d$V$/d$I$ measurements for opposite magnetic fields $B$. Shaded regions mark the direction and range of d.c. current in which the diode is operational. **b,** Rectification measurements in **a**, where the top panel indicates the direction of a d.c. current of magnitude $I$ that is applied in the panels below. From top to bottom, these are $I$ = 30 nA, 32 nA and 34 nA. **d,** Rectification measurements in **c**. From top to bottom they correspond to $I$ = 34 nA, 35 nA and 36 nA. All data was taken at $v$ = -2.9 in device D2.

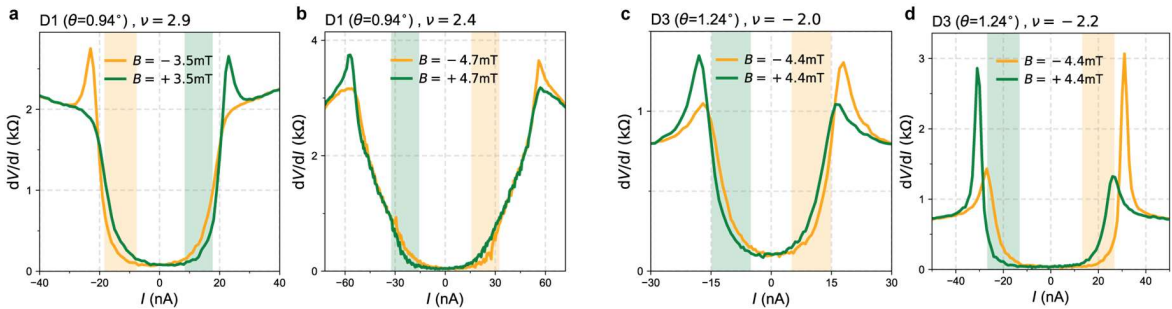

**Fig. S16. Josephson diode in samples D1 and D3.** Differential resistance d$V$/d$I$ vs. d.c. current $I$ at opposite magnetic fields $B$, measured at two different dopings $v$ for each sample. Shaded regions mark the direction and range of the d.c. current in which the superconducting diode is operational.

# I.- Interference patterns of all samples displaying asymmetric oscillations

The interference patterns shown in this section correspond to the raw data (Fig. S17-S20) from where the critical current asymmetry and diode efficiency, plotted in Fig. 5 of the main text, were extracted. This extraction is further explained in Supplementary section K.

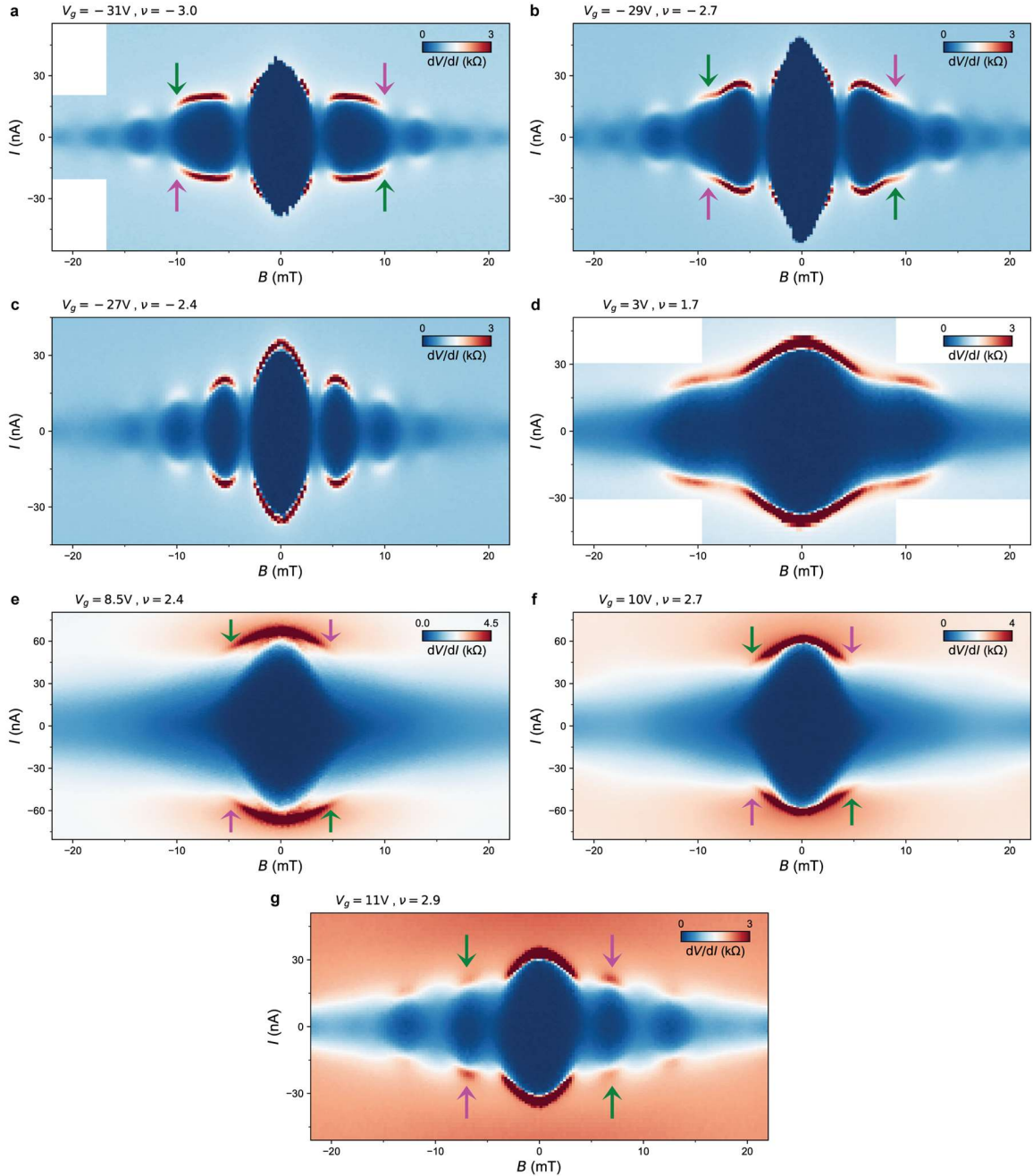

**Fig. S17. Interference patterns measured at $\nu < -2$ and $\nu > 2$ in device D1.** Green and magenta arrows are placed at exact opposite magnetic fields and d.c. currents and indicate regions where the asymmetry is more pronounced. At the edges of the dome, e.g. at dopings $\nu = -2.4$ and $\nu = 1.7$, the asymmetry is lost.

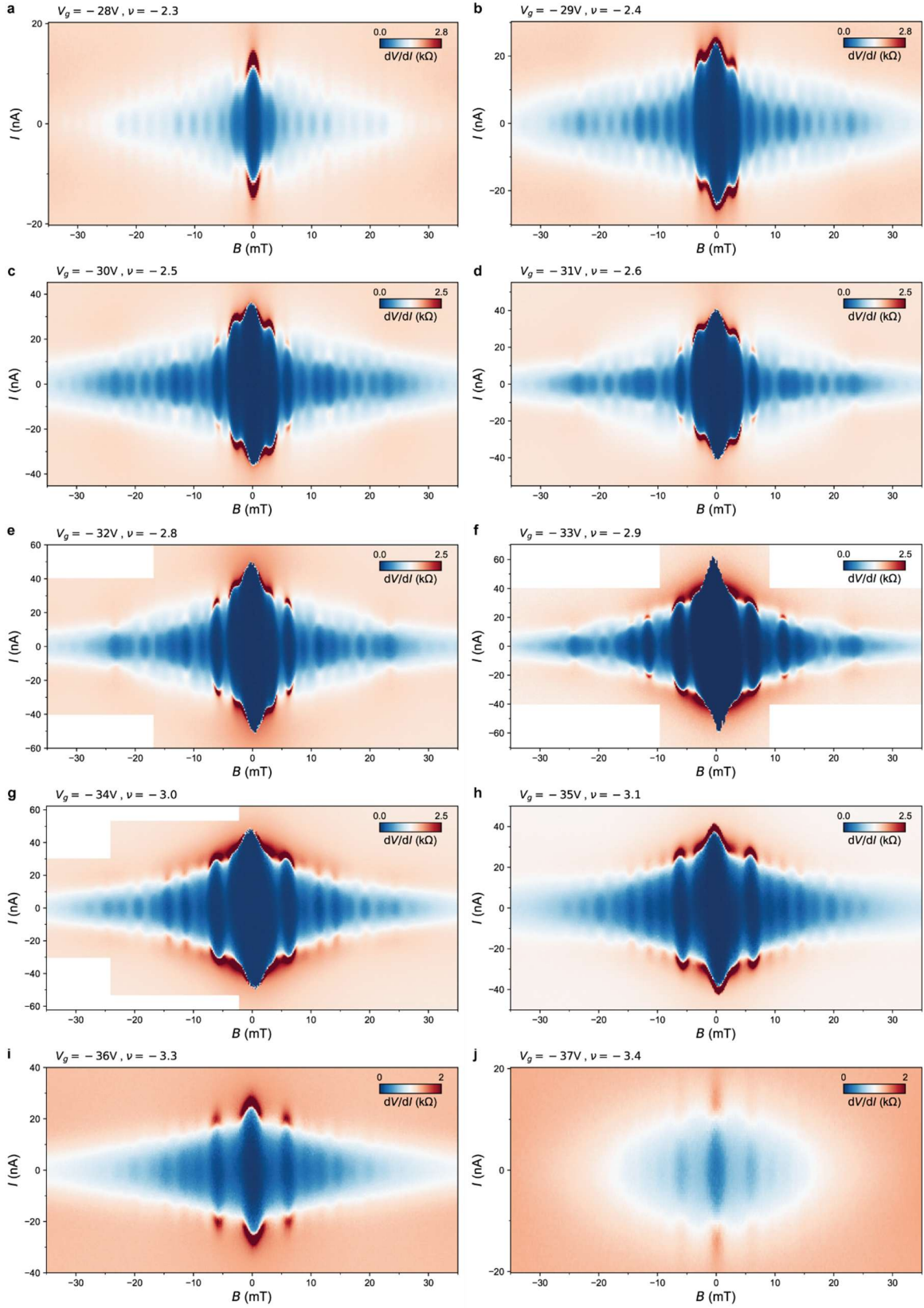

**Fig. S18. Additional interference patterns measured at *v* < -2 in device D2.** Their extracted critical currents correspond to the ones shown in Fig. S22.

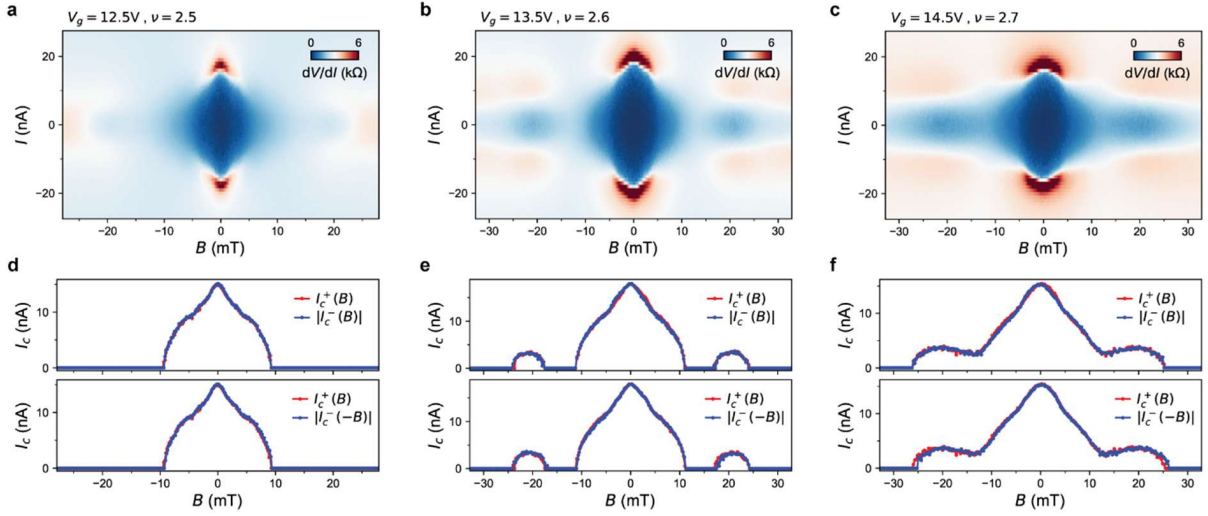

**Fig. S19. Interference patterns measured at $v > 2$ in device D2. a-c,** Interference patterns. **d-f,** Corresponding extracted critical currents for opposite d.c. current directions and plotted for the same and opposite values of magnetic field in the top and bottom panels, respectively. Compared to the hole side of $v = -2$ (see Fig. S18), the asymmetry here is much weaker and harder to detect, given the smaller value of the critical current at these dopings. The relations $I_c^+(B) \neq |I_c^-(B)|$ and $I_c^+(B) = |I_c^-(-B)|$ can only be appreciated between -5 mT and +5 mT.

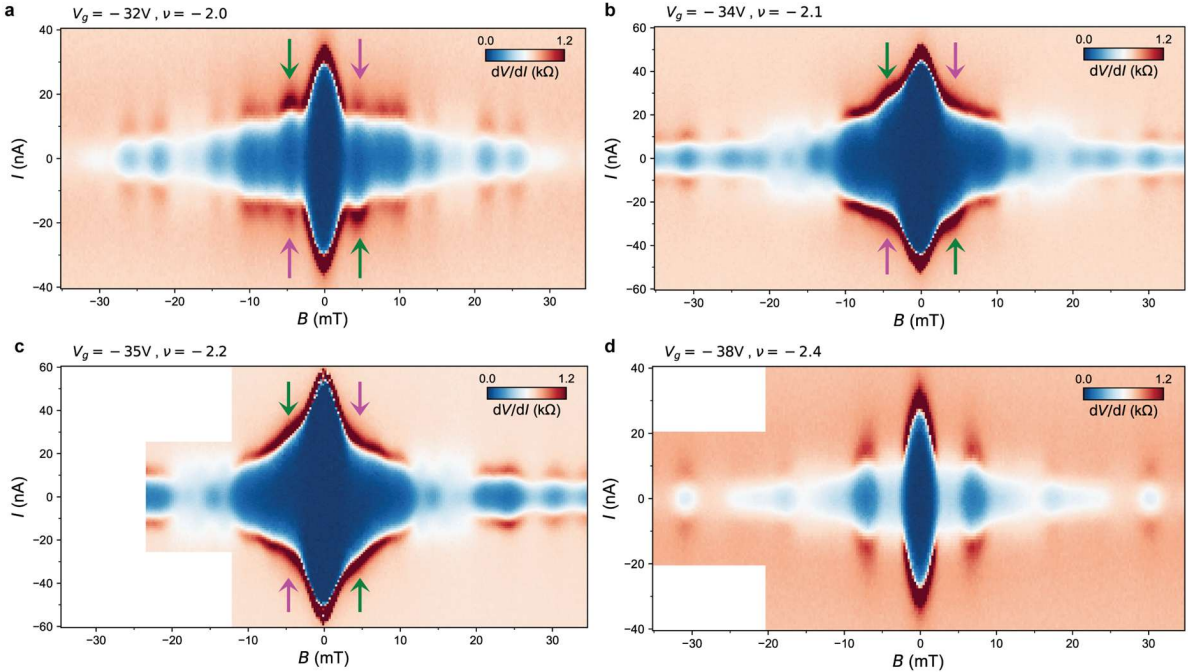

**Fig. S20. Interference patterns measured at $v < -2$ in device D3.** Green and magenta arrows are placed at exact opposite magnetic fields and d.c. currents and indicate regions where the asymmetry is more pronounced. At the edges of the dome, e.g. at doping $v = -2.4$, the asymmetry is lost.

## J.- Temperature dependence of the asymmetric oscillations

The temperature dependence of the interference patterns measured at fixed $v = -2.9$ for device D2, are presented in Fig. S21. There it can be seen that the asymmetric oscillations persist up to $T \sim 0.8$ K, beyond which the supercurrent gets washed out as thermal fluctuations $k_B T$ become comparable to the Josephson energy $E_J = \hbar I_c / 2e$.

We note that the $I_c$ oscillations of the interference patterns where the JDE is found (Fig. 5a of the main text), do not follow an exponential decay proper of a homogeneous supercurrent profile such as in Fig. 1e of the main text. Their slow decay and absence of the first pair of nodes rather point at a strong contribution from edge transport and inhomogeneous supercurrent (see Fig. S25 and Fig. S26). Such SQUID-like inhomogeneous supercurrent density profile, along with the high kinetic inductance of the superconducting TBG [12,13] has been proved to give rise to an asymmetry in the interference patterns [12,14]. If this was the case in our experiment, the inductance associated with the tilt of the interference pattern (Fig. S21e) should increase with temperature and diverge once it reaches the critical temperature [12]. Instead, we find it independent of the temperature.

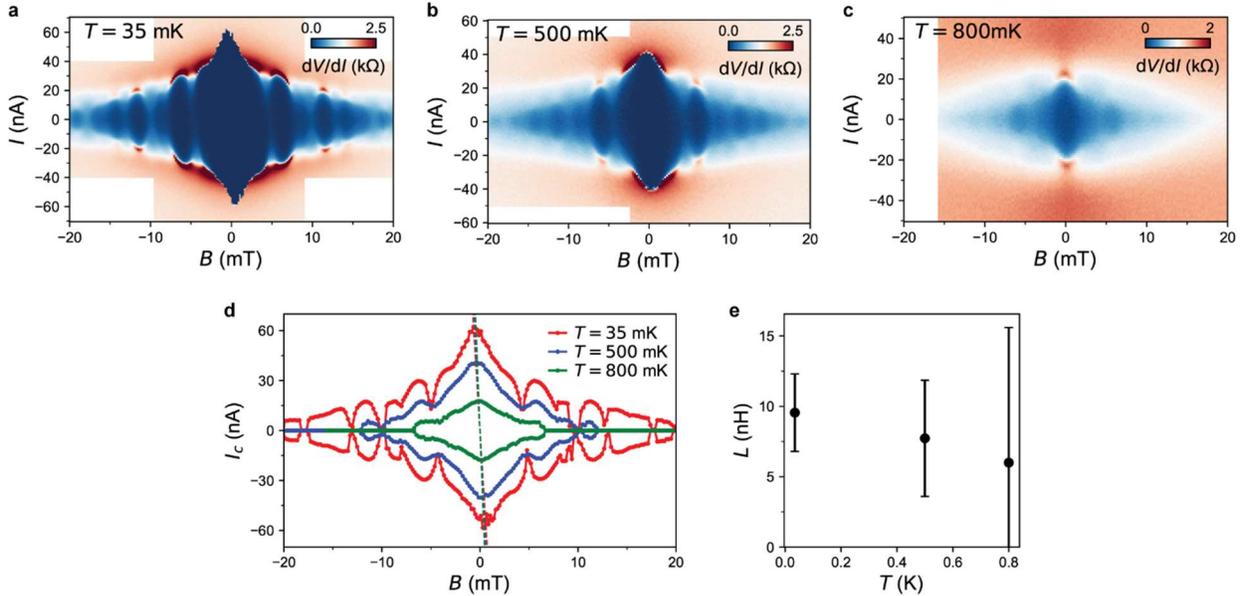

**Fig. S21. Temperature dependence of the asymmetric oscillations. a-c,** Interference patterns measured at fixed $v = -2.9$ for device D2, at different temperatures. **d,** Corresponding extracted critical current. **e,** Extracted inductance $L = \Phi/I$ from the slope of the maxima of the center lobes in the $I_c$ vs $B$ curves in **d**, showing independence with temperature.

### K.- Diode efficiency extraction

In Fig. 5a of the main text, we have shown the relations $I_c^+(B) \neq |I_c^-(B)|$ and $I_c^+(B) = |I_c^-(-B)|$ of the Josephson diode hold at $v = -2.5$ for device D2. In Fig. S22 we show that these relations also hold across the whole $v < -2$ dome. From these traces we can extract the diode efficiency that was shown in Fig. 5d-e of the main text, as we explain in the following.

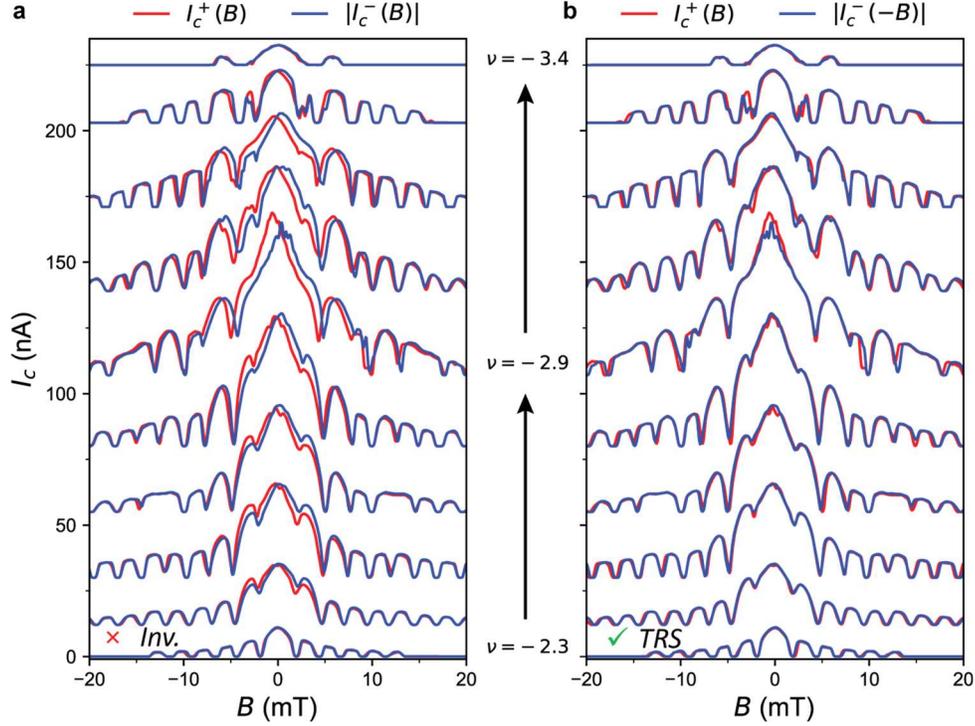

**Fig. S22. Inversion symmetry breaking and TRS conservation at the $v < 2$ dome of D2. a-b,** Critical current $I_c$ as a function of magnetic field $B$ for increasing filling factor $v$, from bottom to top. The curves are each vertically shifted for clarity. **a** shows the mismatch between both directions of the current $I_c^+$ and $I_c^-$, while **b** demonstrates the critical currents are equal upon reversing both current and field directions. All traces were extracted from the raw data in Fig. S18.

In Fig. 5d of the main text, we have calculated the maximum value of the diode efficiency $\eta(B)$ between $-2\Phi_0$ and $2\Phi_0$ in order to compare the JDE between different devices. This is because, at higher fields, the measured dV/dI curves no longer feature a very sharp transition from the superconducting to the normal state. The extracted $I_c$ here is much lower (< 5nA) and the error of extraction is higher. This makes the efficiency $\eta$ to artificially blow up at high fields. In Fig. S23 we show the extracted $\Delta I_c$ and $\eta$ (in absolute value) that were used for Fig. 5d-e of the main text.

We also show an alternative extraction to Fig. 5e by calculating the average value of $\eta(B)$ between -10 mT and +10 mT by doing $<\eta> = \frac{<|\Delta I_c|>}{<I_c^+ + |I_c^-|>}$. The extracted values are shown in Fig. S24. This alternative extraction shows qualitatively an equivalent trend as in Fig. 5e of the main text, with only absolute lower values of $\eta$ in comparison.

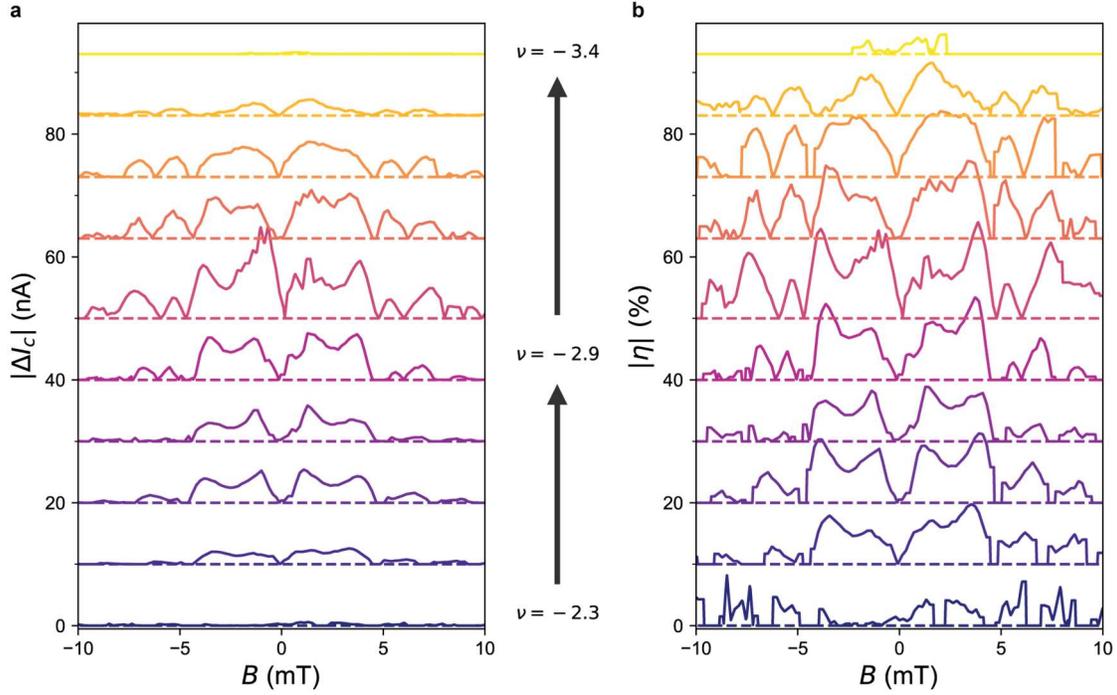

**Fig. S23. Diode efficiency values for device D2. a-b,** Values of $\Delta I_c$ and $\eta$, respectively, represented in absolute value, for different values of filling factor $\nu$ of device D2 from bottom to top. These values were extracted from the traces in Fig. S22. It can be seen, for example in $\nu = -2.3$, that $\eta$ can acquire large values at high fields when $\Delta I_c$ is very low. These (artificially) blow-up $\eta$ values are due to the $I_c$ being very small and thus highly depend on the accuracy of the extraction of $I_c$.

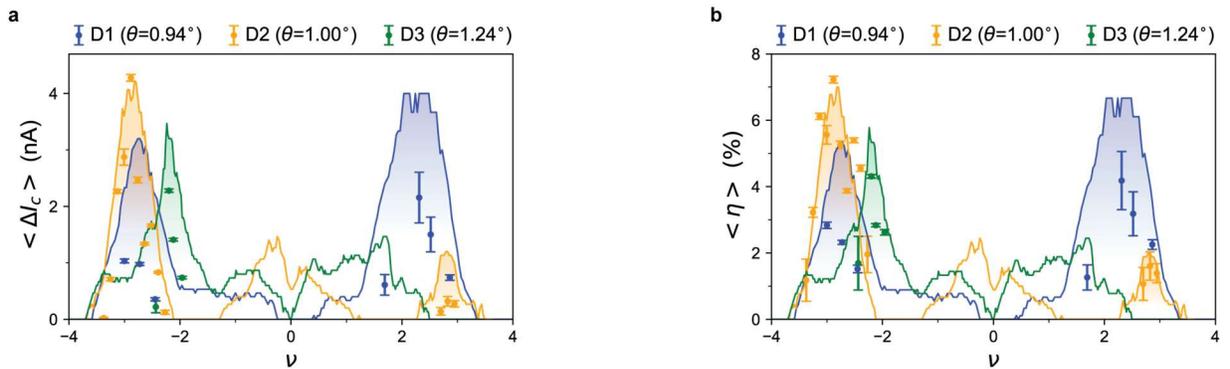

**Fig. S24. Alternative extraction of the asymmetry for devices D1-3. a-b,** Average value of $\Delta I_c(B)$ and $\eta(B)$, respectively, between -10 mT and +10 mT. The values are represented with error bars and as a function of $\nu$ for all junctions near the magic angle D1-3. The line plots correspond to the critical current at zero field (in arbitrary units) as a function of $\nu$. The shaded regions correspond to fillings of the flat-band where a finite asymmetry was recorded.

## L.- Supercurrent carried by edge states

The critical current $I_c$ in a JJ is related to the supercurrent density distribution in real space $J_S$ by the Fourier transform $I_c(\beta) = \left|\int_{-\infty}^{\infty} J_S(x)\, e^{i\beta x}\, dx\right|$, where $x$ is the real space coordinate along the width of the junction and $\beta = 2\pi(L + L_S)B/\Phi_0$ is the normalized magnetic field unit. To retrieve $J_S(x)$ from the experimentally observed $I_c(\beta)$ one can follow the method by Dynes and Fulton [15], where the integral

$$\int_{-\infty}^{\infty} J_S(\beta x)\, e^{i\beta x}\, dx = \int_{-\infty}^{\infty} J_E(x) \cos(\beta x)\, dx + i \int_{-\infty}^{\infty} J_O(\beta) \sin(\beta x)\, dx = I_E(\beta) + i\, I_O(\beta)$$

is divided into a real ($I_E$) and an imaginary ($I_O$) part, which correspond to the even ($J_E$) and odd parts ($J_O$) of the supercurrent distribution respectively. Both $I_E$ and $I_O$ can be computed approximately [16] from the measured $I_c$. Because $I_c(\beta)$ is dominated by $I_E(\beta)$ except at the minima points, one can obtain $I_E(\beta)$ by multiplying $I_c(\beta)$ with a flipping function that switches sign between adjacent lobes. $I_O(\beta)$ is calculated by linearly interpolating between the minima of $I_c(\beta)$ and flipping sign between adjacent lobes. Finally, by computing the inverse Fourier transform of $I_E(\beta) + i\, I_O(\beta)$, the supercurrent density profile $J_S(x)$ is obtained:

$$J_S(x) = \left|\frac{1}{2\pi} \int_{-b/2}^{b/2} \bigl(I_E(\beta) + i\, I_O(\beta)\bigr)\, e^{-i\beta x}\, d\beta\right|,$$

where $b$ is the sampling range of $\beta$, i.e. how far in magnetic field the oscillations were obtained.

By doing this analysis on the interference patterns of our samples where the JDE takes places, we observe that the supercurrent in the junction does not have a uniform profile, but rather skewed and concentrated on the edges of the sample (see Fig. S25a-b). The $I_c$ of the interference patterns from the dispersive bands decay exponentially following the typical Fraunhofer pattern (Fig. S25c-d), although occasionally they can feature some node-lifting (Fig. S25e-f). This results into a slight asymmetry in the supercurrent density profile, where the bulk of the junction, $J_S(x = 0)$, nevertheless carries much more supercurrent than in the case of the interference patterns where the JDE is observed (Fig. S25b). The fact that no JDE is observed in the dispersive bands proves that the asymmetry in the supercurrent profile resulting in node-lifting is not enough no produce a diode effect, and an anomalous or extra term in the current-phase relation is needed. This anomalous term could come from the observed inversion symmetry breaking, which could be correlated with the abundance of edge transport at these fillings compared to others, as just mentioned and evidenced in Fig. S25. We expand on this in the following.

In the main text it is mentioned that the supercurrent is mostly carried by the edges where the JDE is observed. We show this quantitatively in the following. In Fig. S26a we show the extracted supercurrent density profile $J_S(x)$ for all the fillings of the $v < -2$ dome in device D2. To quantify the amount of supercurrent that is carried through the edges and the bulk, we integrate the obtained $J_S(x)$ over the regions shown in Fig. S26b. The results are plotted in Fig. S26c, following the same color-code as in Fig. S26b. The error bars are set by integrating over a 5% bigger/smaller width $x$. Compared to the $I_c$ of the dome measured at zero field (black data), the supercurrents carried by the bulk and the edges follow the same trend, peaking at $v = -2.9$. Furthermore, the left edge of the sample carries significantly more supercurrent than the bulk, while the right edge supports about the same amount. To better visualize this, we plot in Fig. S26d the ratio of the edge and bulk critical currents. Indeed, the right edge $I_c$ (orange) is about one time the bulk, except from the right side of the dome, between $v = -2.3$ and $-2.5$, where it doubles the bulk supercurrent. The left edge $I_c$ (violet) is twice the bulk, reaching three on the right side of the dome.

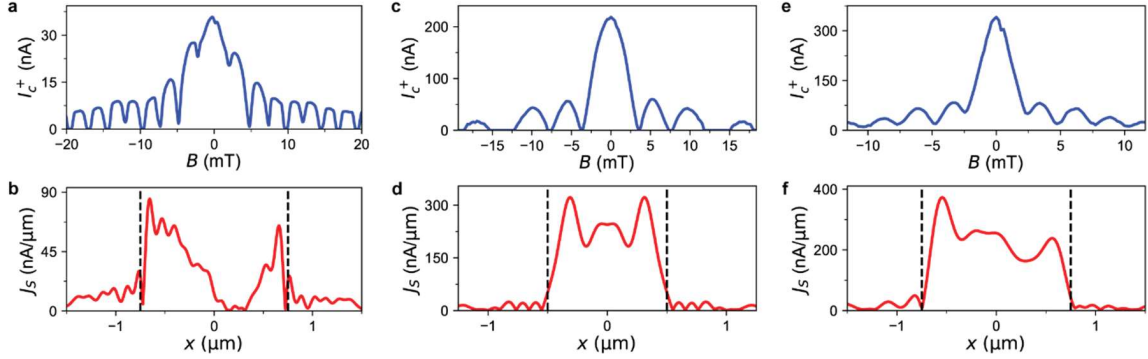

**Fig. S25. Supercurrent density profile and node-lifting. a,** Critical current in one direction of the current, $I_c^+$ vs. magnetic field $B$, extracted from the interference pattern at $v = -2.5$ in D2 from Fig. 5a of the main text. **b,** Corresponding supercurrent density profile $J_S$ vs. the width of the junction $x$. Dashed black lines indicate the physical limit of the junction. The same profile is found for $I_c^-$. **c-d,** Analogous for $v = 8.8$ in D1. **e-f,** Analogous for $v = 7.1$ in D2.

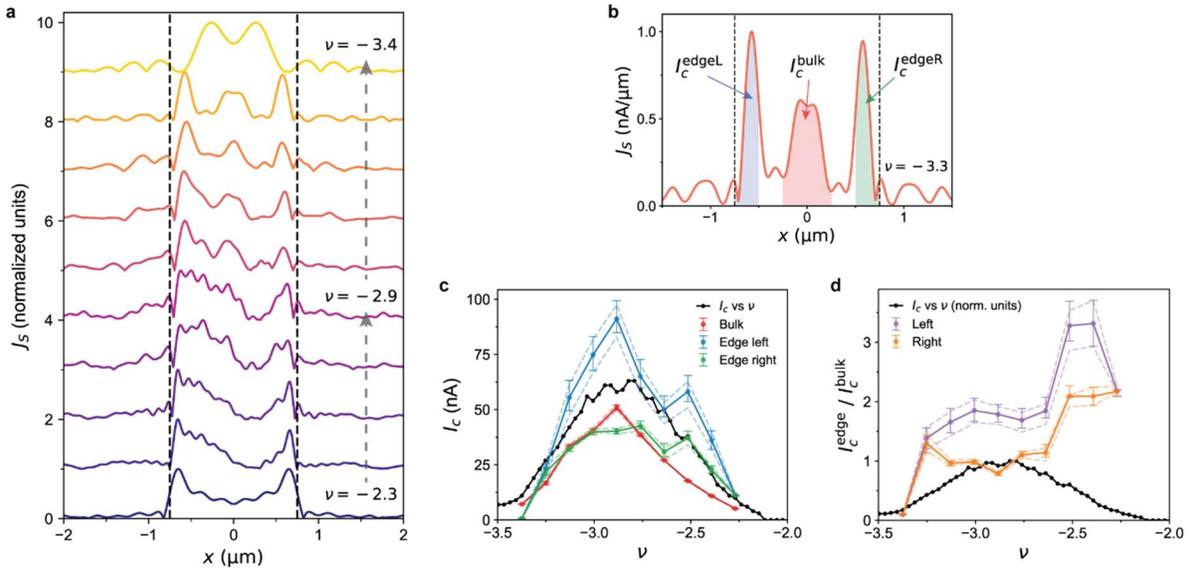

**Fig. S26. Supercurrent density profile in the Josephson diode: Bulk vs Edge. a,** Supercurrent density profile $J_S$ vs. the width of the junction $x$, for different fillings $v$. Each trace is normalized and spaced by one unit. The color-code follows the same trend as in Fig. S23. Dashed black lines indicate the physical limit of the junction. **b,** Normalized $J_S$ vs. $x$ at $v = -3.3$ from **a**. The red, blue and green areas correspond to the critical current in the bulk, left edge and right edge of the sample, respectively. **c,** Critical current $I_c$ vs. $v$. Black data is the extracted $I_c$ at zero field. The $I_c$ of the bulk and the edges are extracted from the traces in **a**, following the example in **b**. **d,** Ratio of the $I_c$ of the edges and the bulk as a function of $v$. Black data is the extracted $I_c$ at zero field, in normalized units. All data corresponds to device D2.

**References:**

[1] D. J. Thoen, B. G. C. Bos, E. A. F. Haalebos, T. M. Klapwijk, J. J. A. Baselmans, and A. Endo, Superconducting NbTin Thin Films With Highly Uniform Properties Over a \varnothing 100 mm Wafer, IEEE Trans. Appl. Supercond. **27**, 1 (2017).

[2] J. G. Kroll et al., Magnetic-Field-Resilient Superconducting Coplanar-Waveguide Resonators for Hybrid Circuit Quantum Electrodynamics Experiments, Phys. Rev. Appl. **11**, 064053 (2019).

[3] P. Dubos, H. Courtois, B. Pannetier, F. K. Wilhelm, A. D. Zaikin, and G. Schön, Josephson critical current in a long mesoscopic S-N-S junction, Phys. Rev. B **63**, 064502 (2001).

[4] M. Tinkham, *Introduction to Superconductivity* (Dover, Mineola, 1996).

[5] H. B. Heersche, P. Jarillo-Herrero, J. B. Oostinga, L. M. K. Vandersypen, and A. F. Morpurgo, Bipolar supercurrent in graphene, Nature **446**, 56 (2007).

[6] J. R. Williams, D. A. Abanin, L. Dicarlo, L. S. Levitov, and C. M. Marcus, Quantum Hall conductance of two-terminal graphene devices, Phys. Rev. B - Condens. Matter Mater. Phys. **80**, (2009).

[7] P. Virtanen, R. P. S. Penttilä, P. Törmä, A. Díez-Carlón, D. K. Efetov, and T. T. Heikkilä, *Superconducting Junctions with Flat Bands*, arXiv:2410.23121.

[8] S. A. Chen and K. T. Law, Ginzburg-Landau Theory of Flat-Band Superconductors with Quantum Metric, Phys. Rev. Lett. **132**, 026002 (2024).

[9] J.-X. Hu, S. A. Chen, and K. T. Law, *Anomalous Coherence Length in Superconductors with Quantum Metric*, https://arxiv.org/abs/2308.05686v5.

[10] R. Bistritzer and A. H. MacDonald, Moiré bands in twisted double-layer graphene, Proc. Natl. Acad. Sci. U. S. A. **108**, 12233 (2011).

[11] D. Hazra et al., Superconducting properties of NbTiN thin films deposited by high-temperature chemical vapor deposition, Phys. Rev. B **97**, 144518 (2018).

[12] E. Portolés, S. Iwakiri, G. Zheng, P. Rickhaus, T. Taniguchi, K. Watanabe, T. Ihn, K. Ensslin, and F. K. de Vries, A tunable monolithic SQUID in twisted bilayer graphene, Nat. Nanotechnol. 2022 1711 **17**, 1159 (2022).

[13] R. Jha, M. Endres, K. Watanabe, T. Taniguchi, M. Banerjee, C. Schönenberger, and P. Karnatak, *Large Tunable Kinetic Inductance in a Twisted Graphene Superconductor*, arXiv:2403.02320.

[14] S. Iwakiri et al., Tunable quantum interferometer for correlated moiré electrons, Nat. Commun. **15**, 1 (2024).

[15] R. C. Dynes and T. A. Fulton, Supercurrent Density Distribution in Josephson Junctions, Phys. Rev. B **3**, 3015 (1971).

[16] S. Hart, H. Ren, T. Wagner, P. Leubner, M. Mühlbauer, C. Brüne, H. Buhmann, L. W. Molenkamp, and A. Yacoby, Induced superconductivity in the quantum spin Hall edge, Nat. Phys. **10**, 638 (2014).